\documentclass[12pt]{article}
\usepackage{amsmath,amsfonts}
\usepackage{algorithmic}
\usepackage{algorithm}
\usepackage{array}
\usepackage[caption=false,font=normalsize,labelfont=sf,textfont=sf]{subfig}
\usepackage{textcomp}
\usepackage{stfloats}
\usepackage{url}
\usepackage{verbatim}
\usepackage{graphicx}
\usepackage{bbm}
\usepackage[numbers]{natbib}

\usepackage{float}
\usepackage[a4paper,margin=1in]{geometry}
\usepackage{setspace}
\newcommand{\be}{\begin{eqnarray}}
\newcommand{\ee}{\end{eqnarray}}
\newcommand{\ba}{\begin{eqnarray*}}
	\newcommand{\ea}{\end{eqnarray*}}
\newcommand{\bei}{\begin{itemize}}
	\newcommand{\beiftnt}{\begin{itemize}\footnotesize}
		\newcommand{\eei}{\end{itemize}}

	\def\mathbold{\boldsymbol}

	\def\bI{\mathbf{I}}

	\def\bx{\mathbf{x}}

	\def\bu{\mathbf{u}}

	\def\bv{\mathbold{v}}

\begin{document}

\title{Recommending Composite Items \\ Using Multi-Level Preference Information:\\ A Joint Interaction Modeling Approach}

\author{Xuan~Bi,~Yaqiong~Wang,~Gediminas~Adomavicius,~Shawn~Curley%
\thanks{X. Bi, G. Adomavicius, and S. Curley are with the Carlson School of Management, University of Minnesota, Minneapolis, MN 55455, USA (e-mail: xbi@umn.edu).}%
\thanks{Y. Wang is with the Leavey School of Business, Santa Clara University, Santa Clara, CA 95053, USA.}}

\date{}


\maketitle

\begin{abstract}
With the advancement of machine learning and artificial intelligence technologies, recommender systems have been increasingly used across a vast variety of platforms to efficiently and effectively match users with items. As application contexts become more diverse and complex, there is a growing need for more sophisticated recommendation techniques. One example is the composite item (for example, fashion outfit) recommendation where multiple levels of user preference information might be available and relevant. In this study, we propose JIMA, a joint interaction modeling approach that uses a single model to take advantage of all data from different levels of granularity and incorporate interactions to learn the complex relationships among lower-order (atomic item) and higher-order (composite item) user preferences as well as domain expertise (e.g., on the stylistic fit). We comprehensively evaluate the proposed method and compare it with advanced baselines through multiple simulation studies as well as with real data in both offline and online settings. The results consistently demonstrate the superior performance of the proposed approach.
\end{abstract}


\clearpage
\section{Introduction}
{R}{ecommender} systems have become ubiquitous across a wide range of fields, such as ecommerce, media consumption (including movies, books, music, news, etc.), social networks, finance, and many others, due to their effectiveness in identifying relevant items or content among numerous choices \citep{adomavicius2005toward,ricci2022recommender}. Traditionally, recommender systems, largely based on collaborative filtering techniques, have focused on recommending individual (or “atomic”) items, such as movies or books, by understanding users’ preferences for these individual items. However, in certain application domains, recommending “composite” items (i.e., combinations of atomic items) represents a very important capability. For illustration, consider a clothing/fashion recommender system, where we want to recommend “outfits” -- combinations of tops (t-shirts, shirts, sweaters) and bottoms (pants, skirts, shorts) -- to users. In such a case, multiple fashion items in a recommended outfit ideally have to match both functionally and stylistically, which may require domain expertise (e.g., on things like style compatibility) beyond individual preferences. 

Another key challenge for such recommender systems is that a given user’s personal preference for a composite item may not directly translate to the user’s personal preferences for the underlying atomic items and vice versa. For example, traditional recommender systems, such as those based on collaborative filtering \citep{funk2006netflix,salakhutdinov2007restricted,bell2007scalable}, could model user preferences for individual items and successfully estimate that a user would love a particular shirt and a pair of pants (individually); however, it is possible that this user would not like the two items together, i.e., as part of the same outfit. Alternatively, traditional approaches could still be used to model such fashion recommender systems by treating entire outfits as recommendable items and learning from users’ outfit preferences, but this would ignore potentially rich and useful information about user preferences for individual tops and bottoms.

Although some solutions have been developed in the existing literature \citep[e.g.,][]{liu2012hi,hu2015collaborative,mcauley2015image,vasileva2018learning} to address the aforementioned problems, several issues remain open for discussion. For example, 
in the fashion recommender systems domain, item compatibility is often learned through processing of apparel images and textual descriptions, which may fail to distinguish compatibility from similarity and unable to capture real fashion expertise (e.g., style, aesthetics, etc.) or personal, subjective preferences for multi-item fit/compatibility. In addition, the incorporation of multi-level user preferences (e.g., preferences at both outfit and individual item levels) is not fully addressed. Part of the challenge lies in accessing and modeling data from multiple sources (e.g., user preferences for both atomic clothing items and outfits), which may require a more specialized recommender system design.
Thus, effective recommendation of composite items that accounts for multi-level user preferences (i.e., revealed preferences for both atomic and composite items) and the relationship between various levels of preferences, e.g., item compatibility, constitutes a promising yet underexplored research setting. This is the focus of our study.

In this article, we use a fashion recommender system as a representative application context and propose a new learning algorithm that is able to provide personalized recommendations of both composite items (e.g., outfits) and atomic items (e.g., tops or bottoms). The proposed method is generalizable to other applications with composite items to generate \textit{compatible} recommendations \citep{han2017learning,vasileva2018learning}. This refers to the recommendation of multiple items that ``fit'' with each other, either subjectively (e.g., based on users' personal preferences for outfit composition) or objectively (e.g., based on style-/fashion-related domain expertise). 
Specifically, we design a multi-task deep learning model based on a joint modeling framework. The proposed method, Joint Interaction Modeling Approach (JIMA), {\em estimates user and item embeddings jointly by taking into account multi-level user preferences},
including atomic item preferences (e.g., subjective user-top and user-bottom ratings), outfit preferences (e.g., subjective user-top-bottom ratings), and domain expertise (e.g., stylistic top-bottom fit ratings).
As part of the proposed approach, we also incorporate {\em two-way and higher-way latent interactions} of user and item embeddings. These interactions can effectively model the compatibility (including personalized, subjective compatibility) between atomic items.

The proposed JIMA method is innovative in several ways. First, it is a systematic deep learning approach that uses a single model to incorporate all data from different levels of granularity. Second, it incorporates multiple interactions to deal with complex relationships between lower-order (atomic item) and higher-order (composite item) preferences. Finally, it allows incorporation of domain expertise (e.g., objective end-user-independent fit ratings). These innovations bring some significant advantages. For example, by incorporating multi-level user preferences, we are able to model and estimate a user’s preference on outfits or individual tops/bottoms with limited preference information, i.e., we may predict a user’s preference on all bottoms with only her ratings on a few tops and outfits. The incorporation of interactions allows the learning of style compatibility without using image or text data. Furthermore, the joint modeling framework can improve accuracy when multiple sources of user preferences are used simultaneously to estimate the user and item embeddings. A comprehensive set of studies, including simulations and both offline and online experiments, was designed to demonstrate the effectiveness of JIMA and illustrate its advantages with respect to a number of baselines.




The rest of this article is organized as follows. Section \ref{sec:lit} discusses existing literature that is related to the proposed work. Section \ref{sec:notation_bgd} introduces the notations and background related to the proposed method. Section \ref{sec:proposed} presents the proposed JIMA method. Sections \ref{sec:sim}, \ref{sec:offline}, and \ref{sec:online} constitute an in-depth evaluation of JIMA using the fashion (i.e., apparel) recommender system as an example application setting. Specifically, in Section \ref{sec:sim}, we conduct comprehensive simulation studies using two-item and three-item outfits and predicting user preferences at various granularities (i.e., preferences for both atomic and composite items), which includes benchmarking against several baselines as well as ablation analyses. In Section \ref{sec:offline}, we collect a set of real rating data for outfits and individual clothing items and provide an offline evaluation of JIMA, again by evaluating predictive performance at various preference granularities, comparing against several baselines, and performing a number of ablation analyses. In Section \ref{sec:online}, we further deploy JIMA online (using a randomized controlled experiment with real users) and evaluate the quality of recommendations that the proposed JIMA method can achieve for human participants as compared to several other recommendation baselines (both algorithmic and non-algorithmic approaches).  The evaluation results consistently point to the effectiveness of JIMA. And finally, Section \ref{sec:discussion} provides concluding remarks and discusses potential directions for future work.

\section{Related Work} \label{sec:lit}

\subsection{Recommender Systems} \label{sec:recsys_lit}

Our study is related to multiple streams of literature in the field of recommender systems. First, a general line of research has proposed a variety of machine learning models that focus on personalized predictions of items that a user may potentially like, interact with, purchase, or consume. Traditionally, recommender systems consist of collaborative filtering \citep{funk2006netflix,salakhutdinov2007restricted,bell2007scalable,bi2017group}, content-based filtering \citep{lang1995newsweeder,mooney2000content}, 
and hybrid recommendations \citep{agarwal2009regression,nguyen2013content}. In particular, due to the presence of contextual and environmental factors, context-aware recommender systems \citep{adomavicius2011context,verbert2012context} and tensor-based recommender systems  \citep{frolov2017tensor,bi2018multilayer,chen2019deep,bi2021tensors,zhang2021dynamic,bi2022improving} are also widely applied. With the fast development of deep learning, the prediction algorithms used for recommender systems are also revolutionalized. This includes, but is not limited to, the use of feedforward neural networks \citep{He2017,guo2017deepfm}, graph neural networks \citep{wu2021self,wu2022graph,gao2023survey}, and transformers \citep{kang2018self,sun2019bert4rec}. Unlike the composite item recommendation in our study, these works generally focus on single item (``atomic'' item) recommendations.

Another closely related stream of literature is on multiple-item recommendations, such as bundle and complementary recommendations. The bundle recommendations \citep{zhu2014bundle,bai2019personalized,deng2020personalized} are motivated by the demand for a personalized set of items. For example, one may receive recommendations for multiple (discounted) video games, TV channels, books, or grocery items. However, in contrast to composite item recommendations (e.g., fashion outfits), bundle recommendations often do not consider the stylistic or subjective compatibility among items. Complementary recommendation refers to recommending an item that is frequently consumed with another \citep{yu2019complementary}. This could include items that are purchased simultaneously \citep{mcauley2015image} or items that have functional compatibility requirements \citep{zhang2018quality}, such as a digital camera and a memory card. However, complementary recommendations do not necessarily refer to compatibility, {\em especially stylistic, aesthetic, or subjective/personalized compatibility}, among the individual items. 

\subsection{Fashion Recommender Systems} \label{sec:fashionrec_lit}

Our study is also related to a growing stream of literature on fashion recommender systems. One active area in fashion recommender systems focuses on single-item recommendations. This includes the search for fashion articles \citep{lasserre2018studio2shop}, and the use of content-based filtering methods to process fashion images and make recommendations for (similar) items \citep{guo2016deep}. One example is using fashion images to predict fashion article sales and make corresponding recommendations to users \citep{bracher2016fashion}. Furthermore, fashion images or photographs can also be used for parsing (i.e., detecting, classifying, or segmenting articles from an image) \citep{yamaguchi2012parsing,liang2016clothes} or for new fashion item design, for example, through convolutional neural networks \citep{kang2017visually} or through generative adversarial networks \citep{kato2019gans}. In summary, this area of research focuses heavily on the extraction, use, or understanding of fashion images; i.e., images constitute key inputs for the corresponding recommendation methods, which is not the focus of the proposed approach. 

In contrast, another body of prior research on fashion recommender systems focuses on stylistic compatibility across fashion items and recommendations \citep{liu2012hi,hu2015collaborative,vasileva2018learning}. For example, this can be done through symbolic word mapping \citep{zhao2017deep}, pair matching \citep{veit2015learning,li2017mining,kuhn2019supporting}, relationships or interactions among users, items, and outfits \citep{li2020hierarchical,lu2021personalized}, and can be applied to session-based recommendations \citep{wu2019session} and outfit generation \citep{han2017learning,chen2019pog}. Similar to the previously discussed stream of literature, many works here rely on fashion images or their latent representations as inputs to their recommendation models.

Our study contributes to the literature on stylistically compatible fashion items. However, the proposed method distinguishes itself from the existing methods in the following ways. 
First, we consider user preferences on both outfits and individual items, and therefore can recommend either outfits, individual items, or both. For example, one user may have only rated \emph{tops} but not other apparel categories, whereas we can still achieve outfit recommendations due to the joint modeling framework. In contrast, many existing methods consider only outfits without the incorporation of user preferences on individual items. 
Second, due to the consideration of user preferences, the proposed method is able to achieve personalized recommendations. In contrast, some existing methods only formulate the co-occurrence of items as an outfit, which to some extent measures popularity rather than individual preferences. Third, the proposed method does not rely on fashion item images or text descriptions and uses only ratings (or views, clicks, purchases, or other interactions in general) to extract user and item embeddings for outfit recommendations. 
Using only rating data leads to fast computation, cost-effective data collection and curation, and efficient machine learning model construction and deployment. 
In contrast, many existing methods require either fashion images or text descriptions in their formulation, which may pose additional requirements or difficulties for data collection and analysis. Lastly, the proposed method incorporates {\em interactions} of latent factors to formulate the complex relationships between atomic and composite preferences (i.e., liking two atomic items separately but not as an outfit). This is another point that has not been thoroughly considered in the existing literature.

\section{Notations and Background} \label{sec:notation_bgd}

\subsection{Notations} \label{notation}

As mentioned earlier, although the proposed approach is applicable in a number of application domains, we use a fashion recommender system as a representative application to present and evaluate our proposed approach.
For notational simplicity, we start from the situation where an outfit consists of two components, namely, \textit{top} and \textit{bottom}. In Section \ref{sec:high_order}, we generalize the proposed method to an arbitrary number of components. Let $Y^{(1)}=\{y^{(1)}_{itb}\}$ be a third-order tensor where each element $y_{itb}$ represents {user} $i$'s preference score on an outfit consisting of {top} $t$ and {bottom} $b$, $i=1,\ldots,N$, $t=1,\ldots,T$, and $b=1,\ldots,B$. Here $N$, $T$, and $B$ are the numbers of unique users, tops, and bottoms, respectively. Further, let $Y^{(2)}=\{y^{(2)}_{it}\}$ and $Y^{(3)}=\{y^{(3)}_{ib}\}$ be matrices with elements representing user $i$'s preference scores on each individual component of the outfit, namely, top $t$ and bottom $b$, respectively. When a user likes a formal jacket and a pair of sporty pants but not the outfit, it is reflected as large $y^{(2)}_{it}$ and $y^{(3)}_{ib}$ values but a small $y_{itb}^{(1)}$ value in our notation. In some cases, we may also have $Y^{(4)}=\{y^{(4)}_{tb}\}$, a matrix of objective fit scores of top $t$ and bottom $b$ acquired via domain knowledge without user preference. (Here the fit scores $y^{(4)}_{tb}$, if available, are usually assigned by independent domain experts who are not users of the recommender system.)

Let $\Omega_1 = \{(i,t,b): y^{(1)}_{itb} \mbox{ is observed}\}$, $\Omega_2 = \{(i,t): y^{(2)}_{it} \mbox{ is observed}\}$, $\Omega_3 = \{(i,b): y^{(3)}_{ib} \mbox{ is observed}\}$, and $\Omega_4 = \{(t,b): y^{(4)}_{tb} \mbox{ is observed}\}$ be the sets of observed values in $Y^{(1)}$, $Y^{(2)}$, $Y^{(3)}$, and $Y^{(4)}$, respectively. In practice, it is common that only a small fraction of elements in $Y^{(1)}$, $Y^{(2)}$, $Y^{(3)}$, and $Y^{(4)}$ are observed, that is, $|\Omega_1|$, $|\Omega_2|$, $|\Omega_3|$, and $|\Omega_4|$ can be small. And it is possible that some of $Y^{(1)}$, $Y^{(2)}$, $Y^{(3)}$, and $Y^{(4)}$ are unobserved entirely. The machine learning goal is to predict the unobserved values in $Y^{(1)}$, $Y^{(2)}$, $Y^{(3)}$, and $Y^{(4)}$. In other words, we predict each user's preference over all individual items and the outfits that they put together to achieve fashionable item recommendation. This can include outfit preference prediction even if some users have only provided ratings about individual items.

\subsection{Background} \label{sec:background}

Traditionally, matrix factorization \citep{funk2006netflix} can be adopted to provide recommendations when we have only one matrix as the information source. For example, for each element $y^{(2)}_{it}$ in matrix $Y^{(2)}$, we can predict it as
\be \label{eq:SVD}
\hat{y}^{(2)}_{it} = \bx_i'\bu_t,
\ee
where $\bx_i = (x_{i1},\ldots,x_{ir})'$ and $\bu_t = (u_{t1},\ldots,u_{tr})'$ are $r$-dimensional latent factors of user $i$ and top $t$, respectively. Here $\bx_i$ and $\bu_t$ can be trained via minimizing a criterion function (e.g., an $L_2$ loss with an $L_2$ penalty) that is defined based on $y^{(2)}_{it}$ and $\hat{y}^{(2)}_{it}$. The optimization process can be achieved through using the gradient descent or alternating least square algorithm. 

Neural collaborative filtering (NCF) \citep{He2017} introduces a general framework that replaces the inner product in \eqref{eq:SVD} with a deep learning structure. That is, they specify
\be \label{eq:NCF}
\hat{y}^{(2)}_{it} = f_{NCF} (\bx_i, \bu_t),
\ee
where $f_{NCF}: \mathbb{R}^{2r} \rightarrow \mathbb{R}$ is a multi-layer neural network (e.g., a feedforward neural network) and uses the concatenation of $\bx_i$ and $\bu_t$ as the input. Here $f_{NCF}$ is characterized by a series of weight matrices and bias vectors, which can be trained via the stochastic gradient descent algorithm.
In fact, the framework in \eqref{eq:NCF} can be generalized to a tensor structure, which is similar to the one developed as neural network based tensor factorization (NTF) \citep{wu2019neural}. For example, for the tensor $Y^{(1)}$, we can have each element $y_{itb}$ predicted as
\be \label{eq:NTF}
\hat{y}^{(1)}_{itb} = f_{NTF} (\bx_i, \bu_t, \bv_b),
\ee
where $\bv_b = (v_{b1},\ldots,v_{br})'$ is an $r$-dimensional latent factor for bottom $b$, and $f_{NTF}: \mathbb{R}^{3r} \rightarrow \mathbb{R}$ is a multi-layer feedforward neural network with the concatenation of $\bx_i$, $\bu_t$, and $\bv_b$ as input.

\section{The Proposed Method} \label{sec:proposed}

\subsection{A Joint Modeling Framework} \label{sec:joint_model}

In this section, we present the proposed method, namely, the Joint Interaction Modeling Approach (JIMA). Recall that in Equation \eqref{eq:NCF} or \eqref{eq:NTF}, information from only one array (e.g., matrix $Y^{(2)}$ or tensor $Y^{(1)}$) is used. The proposed method considers a {\em joint} modeling framework, where latent factors $\bx_i$, $\bu_t$, and $\bv_b$ are estimated using data from $Y^{(1)}$, $Y^{(2)}$, $Y^{(3)}$, and $Y^{(4)}$ simultaneously.

Specifically, each tensor element $y_{itb}^{(1)}$ is predicted as
\be \label{eq:tensor}
\hat{y}^{(1)}_{itb} = f_1 (\bx_i, \bu_t, \bv_b),
\ee
where $f_1$ is a multi-layer feedforward neural network.
The models for predicting elements of $Y^{(2)}$, $Y^{(3)}$, and $Y^{(4)}$ can be constructed similarly, that is,
\be \label{eq:matrices}
\hat{y}^{(2)}_{it} = f_2 (\bx_i, \bu_t), \quad \hat{y}^{(3)}_{ib} = f_3 (\bx_i, \bv_b), \quad \hat{y}^{(4)}_{tb} = f_4 (\bu_t, \bv_b),
\ee
where $f_2$, $f_3$, and $f_4$ are also feedforward neural networks. Notice that, the proposed method has a joint modeling framework in a sense that the same latent factors $\bx_i$, $\bu_t$, and $\bv_b$ are used to construct both \eqref{eq:tensor} and \eqref{eq:matrices}. 
On the other hand, these latent factors are estimated through using data from $Y^{(1)}$, $Y^{(2)}$, $Y^{(3)}$, and $Y^{(4)}$. The joint use of these arrays provides a larger sample size than that of each individual array, and has a potential to achieve better estimation of latent factors (as well as weights and biases in neural networks $f_j$, $j=1,2,3,4$). Here the joint framework guarantees that latent factors can be jointly trained according to more than one factorization, and is applied as long as data from more than one of $Y^{(1)}$, $Y^{(2)}$, $Y^{(3)}$, and $Y^{(4)}$ are available.

The proposed framework can also be generalized to situations where each outfit consists of more than two pieces. For an outfit consisting of $K-1$ pieces, the highest-order tensor essentially has an order $K$, and we theoretically have up to $2^K-K-1$ data sources (after excluding zero- and first-order tensors). For $K=4$, for example, we may have up to 11 data sources. They include a fourth-order tensor denoting user preferences on a three-piece outfit, four third-order tensors denoting either user preferences on two-piece compositions or objective evaluations of three-piece outfits, and six matrices representing user preferences on atomic pieces or objective evaluations of two-piece compositions. The formulation for the proposed joint modeling framework across all $2^K-K-1$ arrays is fully analogous to the situation above, and is hence skipped for notational simplicity.

\subsection{Incorporation of Interactions} \label{sec:interactions}

\begin{figure*}
	\centering \includegraphics[width=0.8\textwidth]{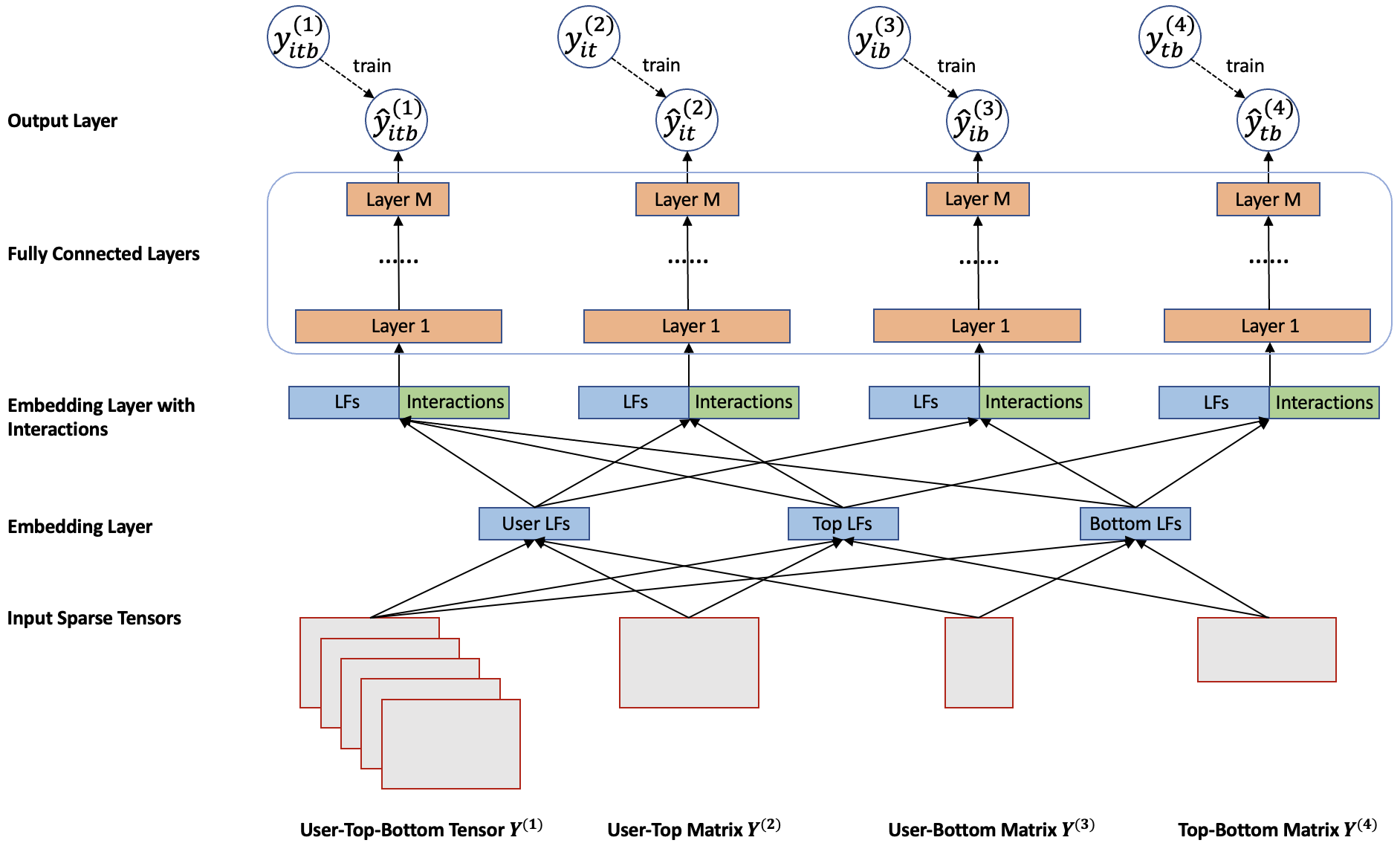}
	\caption{\footnotesize The overview of the proposed method. LF stands for latent factors, and $M$ is the number of fully connected layers in the feedforward neural networks.}
	\label{fig:structure}
\end{figure*}

On one hand, the joint modeling framework provides better estimation of user or item embeddings. On the other hand, however, it may not be sufficient to address the compatibility issue on its own.
Therefore, we also consider two-way and three-way interactions of the latent factors (and higher-way interactions for higher-order tensors), which intend to incorporate the association between preference on individual items and preference on outfits. Let $\bx_i \odot \bu_t$, $\bx_i \odot \bv_b$, and $\bu_t \odot \bv_b$ denote two-way interactions between \emph{user} and \emph{top}, \emph{user} and \emph{bottom}, and \emph{top} and \emph{bottom} latent factors, and $\bx_i \odot \bu_t \odot \bv_b$ denote the three-way interaction, where $\odot$ represents element-wise product (e.g., $\bx_i \odot \bu_t \odot \bv_b = (x_{i1}u_{t1}v_{b1},\ldots,x_{ir}u_{tr}v_{br})'$). It can be seen that, as a special case, these terms allow a user to have positive preference on individual items but negative preference on an outfit 
(when the two-way interaction terms have large negative coefficients) or vice versa. Then, for tensor elements in $Y^{(1)}$, we have
\be \label{eq:tensor_interaction}
\begin{aligned}
&\hat{y}^{(1)}_{itb}= \\
& f_1 (\bx_i, \bu_t, \bv_b, \bx_i \odot \bu_t, \bx_i \odot \bv_b, \bu_t \odot \bv_b, \bx_i \odot \bu_t \odot \bv_b),
\end{aligned}
\ee
where $f_1: \mathbb{R}^{7r} \rightarrow \mathbb{R}$ is a feedforward neural network. Specifically, we concatenate the interaction terms with the latent factors as the input for $f_1()$.
Similarly, the interaction terms for formulating elements of $Y^{(2)}$, $Y^{(3)}$, and $Y^{(4)}$ can be incorporated in
\be \label{eq:matrices_interaction}
\begin{aligned}
\hat{y}^{(2)}_{it} &= f_2 (\bx_i, \bu_t, \bx_i \odot \bu_t),\\
\hat{y}^{(3)}_{ib} &= f_3 (\bx_i, \bv_b, \bx_i \odot \bv_b),\\ 
\hat{y}^{(4)}_{tb} &= f_4 (\bu_t, \bv_b, \bu_t \odot \bv_b),    
\end{aligned}
\ee
where $f_2: \mathbb{R}^{3r} \rightarrow \mathbb{R}$, $f_3: \mathbb{R}^{3r} \rightarrow \mathbb{R}$, and $f_4: \mathbb{R}^{3r} \rightarrow \mathbb{R}$ are feedforward neural networks.
We note that matrices are essentially second-order tensors where only two modes are available. Therefore, three-way interactions are not included in \eqref{eq:matrices_interaction}. Figure \ref{fig:structure} provides a high-level, illustrative overview of the proposed framework.

From a modeling perspective, the optimization of 
all parameters in equations \eqref{eq:tensor_interaction} and \eqref{eq:matrices_interaction} is essentially a multi-task learning problem. The overall criterion function can be written as
\begin{equation} \label{eq:objective}
	\begin{aligned}
		&\frac{1}{|\Omega_1|}  \sum_{ (i,t,b) \in \Omega_1} (y^{(1)}_{itb} - \hat{y}^{(1)}_{itb})^2 +  \frac{1}{|\Omega_2|} \sum_{ (i,t) \in \Omega_2} (y^{(2)}_{it} - \hat{y}^{(2)}_{it})^2\\ &+\frac{1}{|\Omega_3|} \sum_{ (i,b) \in \Omega_3} (y^{(3)}_{ib} - \hat{y}^{(3)}_{ib})^2
		+  \frac{1}{|\Omega_4|} \sum_{ (t,b) \in \Omega_4} (y^{(4)}_{tb} - \hat{y}^{(4)}_{tb})^2 \\
  &+ \sum_{j=1}^4\lambda_j J_j(f_j),
	\end{aligned}
\end{equation}
where $J_j(\cdot)$ and $\lambda_j$ are the $L_2$ penalty function and the associated tuning parameter for regularizing the weights and biases in $f_j$, $j=1,2,3,4$. Each of the four $L_2$ losses in \eqref{eq:objective} can be modified accordingly if the elements of the corresponding tensor or matrix are not continuous. For example, we can have a cross-entropy loss for the tensor or matrices consisting of binary elements. Meanwhile, we note that the proposed model is flexible in terms of data availability. One can simply remove the corresponding loss and penalty functions from \eqref{eq:objective} when some of $\Omega_1$, $\Omega_2$, $\Omega_3$, and $\Omega_4$ are empty. Optimization of \eqref{eq:objective} can be achieved through the use of the Adam algorithm \citep{kingma2014adam}.

\subsection{Generalization to Higher Orders} \label{sec:high_order}

In practice, an outfit may consist of more than two items. This motivates the need to generalize the proposed method to accommodate higher-order tensors. Let $K$ be the order of the highest-order tensor. In other words, we assume that an outfit can consist of up to $(K-1)$ items. In the meantime, an outfit does not necessarily include an item from each category. For example, we may have users' preferences over (hat, shirt, pants), and (shirt, pants, shoes), but not (hat, shoes). Therefore, we further let $L$ be the number of data sources. Here we note that a data source has to include at least two orders (i.e., users' preferences over one-item ``outfits'', or the objective ratings of two-item outfits), and hence $L$ is upper-bounded by $(2^K-K-1)$. For example, for the joint modeling framework introduced in Figure \ref{fig:structure}, we have $K=3$ and $L=4$, denoting tensor order up to 3 and a total of 4 data sources, respectively.

Let $Y$ be the largest possible tensor that has an order $K$ (where $Y$ may not be observed in practice). Let $n_k$ denote the dimension of the $k$th fiber, where $k=1,\ldots,K$. For example, for $Y$ being a tensor of (user, top, bottom), we have top as the second fiber, and $n_2$ denotes the number of unique top items. We also let $Y^{(1)},Y^{(2)},\ldots,Y^{(L)}$ be the $L$ tensors of order up to $K$, whose elements have been (partially) collected. For example, we may have $Y^{(1)}=Y$ be the $K$th-order tensor, denoting users' preferences over outfits consisting of all $(K-1)$ item categories, and $Y^{(L)}$ can be a matrix (i.e., 2nd-order tensor), denoting users' preferences over some single category of atomic items. For $l=1,\ldots,L$, suppose the order of $Y^{(l)}$ is $k_l$, where $k_l \le K$. Here we assume that the $k_l$ fibers of $Y^{(l)}$ correspond to the $l_1$th, $l_2$th, $\ldots$, $l_{k_l}$th fibers of $Y$. For example, if $Y$ has four fibers (user, hat, shirt, pants), then a tensor of (user, shirt, pants) corresponds to the 1st, 3rd, and 4th fibers of $Y$.

Next, we define the latent factors for each individual user and item. Recall that in Section \ref{sec:background}, for $y^{(1)}_{itb}$, we let $\bx_i$, $\bu_t$, and $\bv_b$ be the $r$-dimensional latent factors for user $i$, top $t$, and bottom $b$, respectively. To unify the notation for higher-order tensors, we re-write $\bx_i$, $\bu_t$, and $\bv_b$ as $\bx_{1i}$, $\bx_{2t}$, and $\bx_{3b}$, respectively.
Subsequently, for each of $Y^{(l)}$'s elements $y^{(l)}_{m_1\cdots m_{k_l}}$, we specify an $r$-dimensional latent factor for each of the individuals $m_1, \ldots, m_{k_l}$, that is, $\bx_{l_1m_1}, \ldots, \bx_{l_{k_l}m_{k_l}}$, respectively. Here $l_1,\ldots,l_{k_l}$ denote the corresponding orders of $m_1, \ldots, m_{k_l}$. For example, we may have $l_1=1$ denote the \emph{user} fiber, and $l_2=3$ denote the \emph{shirt} fiber. 

We then apply NTF with interactions similar to Equation \eqref{eq:tensor_interaction}. In other words, we let
\begin{equation} \label{eq:high_order_interaction}
	\begin{aligned}
&\hat{y}^{(l)}_{m_1\cdots m_{k_l}}= \\
&f_l (\bx_{l_1m_1}, \ldots, \bx_{l_{k_l}m_{k_l}}, g_{k_l}(\bx_{l_1m_1}, \ldots, \bx_{l_{k_l}m_{k_l}})),
\end{aligned}
\end{equation}
where $g_{k_l}(\bx_{l_1m_1}, \ldots, \bx_{l_{k_l}m_{k_l}})$ denotes all interactions of $(\bx_{l_1m_1}, \ldots, \bx_{l_{k_l}m_{k_l}})$ as defined by the ``$\odot$'' operation in Section \ref{sec:interactions}, ranging from the second order to the $k_l$th order. And $f_l()$ is a function (e.g., a multi-layer feedforward neural network) that learns $\hat{y}^{(l)}_{m_1\cdots m_{k_l}}$ from $\bx_{l_1m_1}, \ldots, \bx_{l_{k_l}m_{k_l}}$.
It is important to note that, for each dimension $m$ of each fiber $k$, where $k=1,\ldots,K$ and $m=1,\ldots,n_k$, we use the same exact set of latent factors $\bx_{km}$, regardless of which tensors they are in. For instance, for user $i$, the same exact latent factors $\bx_i$ would be used across all learning functions $f_l$ as long as $i$ is involved. This ensures joint modeling. 

Subsequently, the criterion function can be written as
\be \label{eq:general_objective}
		\sum_{l=1}^{L}\frac{1}{|\Omega_l|}  \sum_{ (l_1\cdots l_{k_l}) \in \Omega_l} (y^{(l)}_{m_1\cdots m_{k_l}} - \hat{y}^{(l)}_{m_1\cdots m_{k_l}})^2 + \sum_{l=1}^L\lambda_l J_l(f_l),
\ee
where, similar to Equation \eqref{eq:objective}, $\Omega_l$ denotes the set of observed elements in tensor $Y^{(l)}$, $J_l(\cdot)$ is the regularization function, and $\lambda_j$ is the associated regularization coefficient, $l=1,\ldots,L$.

\section{Simulation Studies} \label{sec:sim}

In this section, we comprehensively simulate the compatibility issue via the use of interactions and demonstrate that the proposed method is able to capture such interactions, leading to improved prediction accuracy. In addition, we 
evaluate the effectiveness of adopting joint modeling.

\subsection{Evaluation of Two-Item Outfits} \label{sec:sim3d}

First, we simulate the case when a user may interact with an outfit consisting of a top and a bottom. We let the tensor size $(N,T,B)$ be $(100,100,100)$, $(50,25,75)$, $(75,100,80)$, and $(150,100,80)$, which simulates different tensor shapes. Here the last three sizes correspond to the cases of bottom-majority, top-majority, and user-majority, respectively. And the third and fourth sizes are the same, except the number of users is doubled in the fourth; the purpose of this is to illustrate how the data size may change the prediction results.

We then simulate shared latent factors which will be jointly modeled. Specifically, we generate each $\bx_i$ ($i=1,\ldots,N$), $\bu_t$ ($t=1,\ldots,T$), and $\bv_b$ ($b=1,\ldots,B$) independently from a multivariate standard normal distribution $N(\mathbf{0},\bI_r)$, where the number of latent factors $r$ is set as 5, and $\bI_r$ denotes an $r$-dimensional identity matrix. To simulate interactions as part of the ground truth, we let
\ba
y^{(1)}_{itb} &=& \bx_i'\bu_t + \bx_i'\bv_b + 2  \bu_t'\bv_b + e^{(0)}_{itb}\\
y^{(2)}_{it} &=& \bx_i'\bu_t + e^{(1)}_{it}\\
y^{(3)}_{ib} &=& \bx_i'\bv_b + e^{(2)}_{ib}\\
y^{(4)}_{tb} &=& \bu_t'\bv_b,
\ea
where all $e^{(k)}$ are simulated noise terms that independently follow $N(0,0.1^2)$. The inner products $\bx_i' \bu_t$, $\bx_i'\bv_b$, and $\bu_t'\bv_b$ intend to capture the pairwise similarity between user, top, and bottom latent factors. For $y_{itb}^{(1)}$, we incorporate interaction term $2  \bu_t'\bv_b$ in order to simulate the compatibility issue. In other words, some user $i$ may like top $t$ (e.g., a formal suit) and bottom $b$ (e.g., sporty pants) separately, but not as an outfit because of the poor match. This would be reflected as a positive $\bx_i'\bu_t$, a positive $\bx_i'\bv_b$, but a very negative $2  \bu_t'\bv_b$. Here we do not inject noise for $y^{(4)}_{tb}$ as they are considered the objective matching score of top $t$ and bottom $b$.

We simulate $(user,top,bottom)$-tensor $Y^{(1)}$, $(user,top)$-matrix $Y^{(2)}$, $(user,bottom)$-matrix $Y^{(3)}$, and $(top,bottom)$-matrix $Y^{(4)}$ entirely. For each method, our training data consist of a concatenation of $20\%$ random elements in each of $Y^{(1)}$, $Y^{(2)}$, $Y^{(3)}$, and $Y^{(4)}$. And the rest $80\%$ of elements in $Y^{(1)}$, $Y^{(2)}$, $Y^{(3)}$, and $Y^{(4)}$ are concatenated and used as the test set. 

The proposed method, JIMA, is compared with several illustrative baseline approaches: global mean imputation (GMI; predicting every value as the mean of all observed values in the corresponding tensor); two tensor methods for outfit recommendations, including the classical tensor CP decomposition \citep[TF (CPD);][]{carroll1970analysis} and neural tensor factorization \citep[NTF;][]{wu2019neural};\footnote{The original NTF was designed for dynamic data. We use the neural tensor factorization framework for comparison, but without the LSTM time encoder.} two matrix methods for single item recommendations (i.e., \emph{tops} or \emph{bottoms}), including matrix factorization \citep{funk2006netflix} and neural collaborative filtering \citep[NCF;][]{He2017}. Note that none of these approaches utilize multiple data sources or considers interactions of latent factors. MF and NCF only use information from $Y^{(2)}$, $Y^{(3)}$, and $Y^{(4)}$, and TF (CPD) and NTF only use information from $Y^{(1)}$. Here MF and TF are traditional factorization-based methods, while NCF and NTF are deep learning methods. For all NCFs and NTFs, we consider a multi-layer perceptron with 4 layers, consisting of 64, 32, 16, and 8 neurons, respectively.

As importantly, for the proposed method, we also consider the ablated versions of JIMA, by removing either the interaction-based or the joint-modeling-based component.
For notational simplicity, we refer to $Y^{(1)}$, $Y^{(2)}$, $Y^{(3)}$, and $Y^{(4)}$ as the utb-tensor, ut-matrix, ub-matrix, and tb-matrix, respectively. Also, let NF (as in ``Neural Factorization'') denote either NTF or NCF, whenever the context is clear. For example, we use NF$_{\mbox{utb}}$ to denote the NTF of utb-tensor, and NF$_{\mbox{ut}}$ to denote the NCF of ut-matrix.
We use NF$^{\mbox{x}}$ to denote the proposed method with interactions only (without any joint modeling). For example, we use NF$^{\mbox{x}}_{\mbox{utb}}$ to denote the NTF of utb-tensor with all two-way and three-way interactions, as defined in Equation \ref{eq:tensor_interaction}.
And we use ``$\oplus$'' to denote the proposed method with joint modeling of the specified tensors and/or matrices. For example, NF$_{\mbox{utb}\oplus\mbox{ut}}$ denotes the joint modeling of utb-tensor NTF and ut-matrix NCF (without interactions).  Based on this notation, NF$^{\mbox{x}}_{\mbox{utb}\oplus\mbox{ut}\oplus\mbox{ub}\oplus\mbox{tb}}$ corresponds to the full JIMA as it was described in Section \ref{sec:proposed}, with both joint modeling and interactions.
An analogous naming convention is also adopted in Section \ref{sec:sim4d}.

The predictive performance results, measured using the root mean square error (RMSE) metric, corresponding to $(N,T,B)$ being $(100,100,100)$ are summarized in Table \ref{Sim100-100-100}. Table \ref{Sim100-100-100} also indicates the sources of data that each method uses (i.e., empty cells correspond to information sources that are not used). Recall that the existing methods (e.g., NTF, NCF) are able to use information from one data source at a time, and hence their results on other three datasets are not available. Due to page limits, the additional results corresponding to $(N,T,B)$ being $(50,25,75)$, $(75,100,80)$, and $(150,100,80)$ are summarized in Tables \ref{Sim50-25-75}, \ref{Sim75-100-80}, and \ref{Sim150-100-80} of Appendix, respectively. The analogous predictive performance results, measured using the mean absolute error (MAEs) metric, as well as the computational runtimes are also provided in Appendix (i.e., Tables \ref{Sim100-100-100-mae}, \ref{Sim50-25-75-mae}, \ref{Sim75-100-80-mae}, and \ref{Sim150-100-80-mae}). The entire process, from simulating synthetic data to evaluating the results, is replicated 50 times.

It can be seen that, across all $(N,T,B)$ shapes, the proposed approach demonstrates excellent performance. In particular, the proposed method (using all information) leads to the best performance overall. Specifically, its advantages are reflected in three aspects.
First, the proposed method outperforms existing methods including TF, MF, NTF, and NCF. For example, in Table \ref{Sim100-100-100}, the RMSE of the proposed method for the utb-tensor is 0.198, whereas the best performance among existing methods has RMSE of 0.421. Moreover, RMSEs of the proposed method for the three matrix sources are between 0.155 and 0.216, whereas RMSEs of the existing methods are between 2.091 and 2.331.
Second, the proposed method significantly improves the ablated versions that do not use interactions, where the RMSE improvement ranges from 47.0\% to 64.4\%. This indicates the crucial value of interactions in prediction.
Third, the proposed method outperforms its NF$^{\mbox{x}}$ versions (i.e., its ablated versions without joint modeling).
On one hand, the proposed method has similar performance as NF$^{\mbox{x}}$ on the utb-tensor; on the other hand, it is up to 93.1\% better than NF$^{\mbox{x}}$ that do not use the joint modeling on the three matrices. The results in the Appendix show consistent findings.

Interestingly, by comparing JIMA to NTF (= NF$_{utb}$) and to various ablated versions of JIMA on the 3-dimensional utb-tensor data, we find that the performance of the proposed method benefits more from the inclusion of interactions. However, by comparing JIMA to NCF and to various ablated versions of JIMA on the different 2-dimensional data matrices (i.e., ut-matrix, ub-matrix, and tb-matrix), we find that the more substantial performance benefits come from the joint modeling capability. One explanation is that the tensor has a larger sample size and, thus, is more ``saturated'' in terms of model training for composite-item preference prediction. Therefore, the inclusion of more lower-level (i.e., atomic-item) preference information from the three matrices through joint modeling does not bring much improvement, whereas the enhancement of the model structure (i.e., the incorporation of potential interactions between multiple atomic elements of composite items) is more critical, especially in this specific situation where interactions between items in an outfit are strongly present. In contrast, in each of the three 2-dimensional matrices, modeling the atomic-item preference information benefits from being able to bring additional sources of information using the joint modeling approach, whereas formulating interactions on top of atomic-item preference data generally does not improve the performance of the canonical matrix factorization (and can add to the risk of overfitting).

The observations above demonstrate the effectiveness of combining joint modeling and interactions. In fact, the capability to provide predictions across all data sources simultaneously brings an additional advantage. That is, after model training,
the proposed method enables the recommendation of outfits to a user even if the user has never rated any outfits. This is because joint modeling allows user and item embeddings to be learned through other data sources. 
In contrast, tensor factorization methods require that the user has rated some outfits and that all atomic items appearing in outfits are included in the dataset so their embeddings can be learned.

\begin{table*}
\begin{tiny}
\caption{Simulated outfit recommendation accuracy comparison between the proposed methods (JIMA) and the competing methods with $N=100$, $T=100$, and $B=100$ using RMSE (with standard deviation in parentheses). The best performance is highlighted in bold.}
\vspace{-4mm}
\begin{center}
\begin{tabular}{cccccc}
\hline
\hline
& & \multicolumn{4}{c}{Data sources for measuring prediction performance} \\
\hline
Model Type & Model Name & utb-tensor & ut-matrix & ub-matrix & tb-matrix \\
\hline
Global Mean Imputation & GMI & 5.505 (0.220) & 2.264 (0.117) & 2.259 (0.103) & 2.240 (0.094) \\
\hline
Tensor/Matrix Factorization & TF (CPD) & 3.120 (0.121) &  &  &  \\
& MF &  & 2.110 (0.107) &  &  \\
& MF &  &  & 2.091 (0.193) &  \\
& MF &  &  &  & 2.100 (0.067) \\
\hline
Deep Learning Benchmarks & NTF (=NF$_{\mbox{utb}}$) & 0.421 (0.021) &  &  &  \\
& NCF (=NF$_{\mbox{ut}}$) &  & 2.331 (0.176) &  &  \\
& NCF (=NF$_{\mbox{ub}}$) &  &  & 2.189 (0.005) &  \\
& NCF (=NF$_{\mbox{tb}}$) &  &  &  & 2.276 (0.361) \\
\hline
\hline \\[-0.8em]
JIMA & NF$^{\mbox{x}}_{\mbox{utb}\oplus\mbox{ut}\oplus\mbox{ub}\oplus\mbox{tb}}$ & 0.198 (0.030) & {\bf 0.216 (0.046)} & {\bf 0.213 (0.044)} & {\bf 0.155 (0.058)} \\
\hline
\hline\\[-0.8em]
Ablated JIMA (interactions only) & NF$^{\mbox{x}}_{\mbox{utb}}$ & {\bf 0.173 (0.011)} &  &  &  \\[0.5em]
\vspace{1mm}
& NF$^{\mbox{x}}_{\mbox{ut}}$ &  & 2.237 (0.019) &  &  \\[0.2em]
\vspace{1mm}
& NF$^{\mbox{x}}_{\mbox{ub}}$ &  &  & 2.263 (0.237) &  \\[0.2em]
& NF$^{\mbox{x}}_{\mbox{tb}}$ &  &  &  & 2.253 (0.186) \\[0.2em]
\hline
Ablated JIMA (joint modeling only) & NF$_{\mbox{utb}\oplus\mbox{ut}}$ & 0.426 (0.005) & 0.469 (0.017) &  &  \\
& NF$_{\mbox{utb}\oplus\mbox{ut}\oplus\mbox{ub}}$ & 0.415 (0.032) & 0.456 (0.025) & 0.402 (0.022) &  \\
& NF$_{\mbox{utb}\oplus\mbox{ut}\oplus\mbox{ub}\oplus\mbox{tb}}$ & 0.431 (0.015) & 0.468 (0.088) & 0.458 (0.056) & 0.435 (0.055) \\
\hline
\hline
\end{tabular}
\end{center}
\label{Sim100-100-100}
\end{tiny}
\end{table*}

\subsection{Evaluation of Three-Item Outfits} \label{sec:sim4d}

Here we simulate the case where users interact with more complex composite items consisting of up to three atomic items (e.g., a top, a bottom, and a hat). This requires tensors up to an order of $K=4$. Following commonly seen fashion outfit conventions, we assume that a user can rate: a top, a bottom, or a hat, individually; an outfit that consists of a top and a bottom; or an outfit that consists of a top, a bottom, and a hat. Following our discussion in Section \ref{sec:high_order}, this corresponds to $L=5$ tensors:  (\emph{user}, \emph{top}), (\emph{user}, \emph{bottom}), (\emph{user}, \emph{hat}), (\emph{user}, \emph{top}, \emph{bottom}), and (\emph{user}, \emph{top}, \emph{bottom}, \emph{hat}). In other words, in this example experiment, we do not allow tensors consisting of other combinations (e.g., (\emph{user}, \emph{bottom}, \emph{hat})).

Specifically, let $(n_1,n_2,n_3,n_4)$ denote the tensor dimensions corresponding to \emph{user}, \emph{top}, \emph{bottom}, and \emph{hat}, respectively. We simulate five different combination shapes, $(100, 20, 15, 10)$, $(100, 10, 15, 20)$, $(100, 15, 20, 15)$, $(150,20, 20, 20)$, and $(150,20,30,20)$. The first three combination shapes imitate the situations when we have more tops, hats, bottoms, respectively, and the last two shapes are designed to evaluate the cases when the sample size is larger.

Following the notation of Section \ref{sec:high_order}, we generate each $\bx_{km}$ following a multivariate standard normal distribution $N(\mathbf{0},\bI_r)$ with the number of latent factors $r=5$. Here $k=1,2,3,4$ represents the fibers for \emph{user}, \emph{top}, \emph{bottom}, \emph{hat}, respectively, and $m=1,\ldots,n_k$ represents each individual subject along the $k$th fiber. For example, $\bx_{2,8}$ denotes the latent factors of the 8th \emph{top} item. Then the elements of the tensors $Y^{(l)}$, $l=1,\ldots,5$, are generated as follows:
\ba
y^{(1)}_{itbh} &=& \bx_{1i}'\bx_{2t} + \bx_{1i}'\bx_{3b} + \bx_{1i}'\bx_{4h} \\
&&+ 2  \bx_{2t}'\bx_{3b} + 2  \bx_{2t}'\bx_{4h} + 2  \bx_{3b}'\bx_{4h} + e^{(5)}_{itbh}\\
y^{(2)}_{itb} &=& \bx_{1i}'\bx_{2t} + \bx_{1i}'\bx_{3b} + 2  \bx_{2t}'\bx_{3b} + e^{(4)}_{itb}\\
y^{(3)}_{it} &=& \bx_{1i}'\bx_{2t} + e^{(1)}_{it}\\
y^{(4)}_{ib} &=&  \bx_{1i}'\bx_{3b}  + e^{(2)}_{ib}\\
y^{(5)}_{ih} &=&  \bx_{1i}'\bx_{4h} + e^{(3)}_{ib},
\ea
where all $e^{(l)}$ are simulated noise terms that independently follow $N(0,0.1^2)$ for $l=1,\ldots,5$. 

The proposed JIMA approach is compared with the same set of competing methods as in Section \ref{sec:sim3d}. Similarly, 20\% of elements from each tensor are combined as the training data and the rest is used for testing. We replicate each setting 50 times and consider ablated versions of the proposed method. 

The simulation results for the largest shape $(150,20,30,20)$ are provided in Table \ref{Sim4d150-20-30-20}, using the same notational conventions as in Table \ref{Sim100-100-100}. Similar to the previous simulation, the source of data that each method uses, including the sources used by the competing methods and the ablated versions of JIMA, are also reported in Table \ref{Sim4d150-20-30-20} (i.e., via non-empty cells). Due to page limits, other simulation results are provided in Tables \ref{Sim4d100-20-15-10}, \ref{Sim4d100-10-15-20}, \ref{Sim4d100-15-20-15}, and \ref{Sim4d150-20-20-20} in the Appendix, corresponding to the shapes $(100, 20, 15, 10)$, $(100, 10, 15, 20)$, $(100, 15, 20, 15)$, and $(150,20, 20, 20)$, respectively. The corresponding MAE and computational runtimes are also provided in Appendix (i.e., Tables \ref{Sim4d150-20-30-20-mae}, \ref{Sim4d100-20-15-10-mae}, \ref{Sim4d100-10-15-20-mae}, \ref{Sim4d100-15-20-15-mae}, and \ref{Sim4d150-20-20-20-mae}).

Again, it can be seen that JIMA achieves the best performance on almost all data sources, across all settings, and compared with all competing methods. Specifically, for the highest-order tensor (i.e., utbh-tensor), incorporating interactions into NTF significantly reduces prediction error (as measured by RMSE); and the additional joint modeling provides no further improvement to JIMA. For lower-order tensors (i.e., ut-matrix), incorporating interactions offers less of the benefit in terms of RMSE reduction, whereas the incorporation of joint modeling leads to a much greater improvement. As discussed in Section \ref{sec:sim3d}, one possible reason is that the highest-order tensor contains the most observations, and hence is relatively self-contained for embedding learning; consequently, adding preference information via joint modeling may not necessarily yield further improvement. On the other hand, the incorporation of interactions meaningfully alters the model formulation by accounting for the compatibility among items. This leads to the greatest performance improvement for higher-order tensors. In contrast, the lower-order tensors (with smaller numbers of observations) again benefit more drastically from joint modeling due to their access to additional preference data, whereas the ability to incorporate interactions offers limited additional value for the lower-level (i.e., atomic-item) preference modeling.

In summary, evaluation studies in both sections \ref{sec:sim3d} and \ref{sec:sim4d} (and corresponding appendices) underscore the value of both components, i.e., joint modeling and interaction modeling, in the proposed design that facilitates accurate multi-level preference predictions using a single, unified model.

\begin{table*}
\begin{tiny}
\caption{Simulated outfit recommendation accuracy comparison between the proposed methods (JIMA) and the competing methods with shape $(150, 20, 30, 20)$ using RMSE (with standard deviation in parentheses). The best performance is highlighted in bold.}
\vspace{-4mm}
\begin{center}
\begin{tabular}{ccccccc}
\hline
\hline
& & \multicolumn{5}{c}{Data sources for measuring prediction performance} \\
\hline
Type & Model Name & utbh-tensor & utb-tensor & ut-matrix & ub-matrix & uh-matrix \\
\hline
Global Mean Imputation & GMI 
& 8.692 (0.680) & 5.586 (0.472) & 2.277 (0.180) & 2.269 (0.142) & 2.216 (0.178) \\
\hline
Tensor/Matrix & TF (CPD) 
& 4.115 (0.397) &  &  &  &  \\
Factorization & TF (CPD) 
&  & 3.280 (0.245) &  &  &  \\
& MF 
&  &  & 2.269 (0.186) &  &  \\
& MF 
&  &  &  & 2.195 (0.139) &  \\
& MF 
&  &  &  &  & 2.123 (0.136) \\
\hline
\vspace{1mm}
Deep Learning & NTF (=NF$_{\mbox{utbh}}$) 
& 0.434 (0.042) &  &  &  &  \\
\vspace{1mm}
Benchmarks & NTF (=NF$_{\mbox{utb}}$) 
&  & 0.865 (0.165) &  &  &  \\
\vspace{1mm}
& NCF (=NF$_{\mbox{ut}}$) 
&  &  & 2.681 (0.260) &  &  \\
\vspace{1mm}
& NCF (=NF$_{\mbox{ub}}$) 
&  &  &  & 2.576 (0.197) &  \\
\vspace{1mm}
& NCF (=NF$_{\mbox{uh}}$) 
&  &  &  &  & 2.618 (0.258) \\
\hline
\hline\\[-0.8em]
JIMA & NF$^{\mbox{x}}_{\mbox{utbh}\oplus\mbox{utb}\oplus\mbox{ut}\oplus\mbox{ub}\oplus\mbox{uh}}$ 
& 0.209 (0.017) & \textbf{0.240 (0.040)} & \textbf{0.309 (0.091)} & \textbf{0.267 (0.055)} & \textbf{0.298 (0.088)} \\
\hline
\hline\\[-0.8em]
Ablated JIMA & NF$^{\mbox{x}}_{\mbox{utbh}}$ 
& \textbf{0.170 (0.011)} &  &  &  &  \\[0.5em]
(interactions only) & NF$^{\mbox{x}}_{\mbox{utb}}$ 
&  & 0.424 (0.226) &  &  &  \\[0.5em]
\vspace{1mm}
& NF$^{\mbox{x}}_{\mbox{ut}}$ 
&  &  & 2.699 (0.267) &  &  \\[0.2em]
\vspace{1mm}
& NF$^{\mbox{x}}_{\mbox{ub}}$ 
&  &  &  & 2.561 (0.236) &  \\[0.2em]
\vspace{1mm}
& NF$^{\mbox{x}}_{\mbox{uh}}$ 
&  &  &  &  & 2.629 (0.266) \\
\hline
\vspace{1mm}
Ablated JIMA & NF$_{\mbox{utbh}\oplus\mbox{utb}}$ 
& 0.450 (0.045) & 0.613 (0.088) &  &  &  \\
\vspace{1mm}
(joint modeling only) & NF$_{\mbox{utbh}\oplus\mbox{utb}\oplus\mbox{ut}}$ 
& 0.477 (0.044) & 0.621 (0.064) & 0.601 (0.083) &  &  \\
\vspace{1mm}
& NF$_{\mbox{utbh}\oplus\mbox{utb}\oplus\mbox{ut}\oplus\mbox{ub}}$ 
& 0.467 (0.043) & 0.627 (0.073) & 0.617 (0.071) & 0.543 (0.070) &  \\
\vspace{1mm}
& NF$_{\mbox{utbh}\oplus\mbox{utb}\oplus\mbox{ut}\oplus\mbox{ub}\oplus\mbox{uh}}$ 
& 0.475 (0.049) & 0.614 (0.064) & 0.624 (0.087) & 0.623 (0.368) & 0.606 (0.112) \\
\hline
\hline
\end{tabular}
\end{center}
\label{Sim4d150-20-30-20}
\end{tiny}
\end{table*}

\section{Offline Experiments with Real Data} \label{sec:offline}


In the previous section, we use simulation experiments to demonstrate the ability of the proposed method to capture the existing joint modeling and interaction patterns in multi-level preferences. In practice, however, the degree to which these patterns exist and can be learned to enhance preference prediction is unknown. In this section, we demonstrate the capability of the proposed method in such a scenario.

There is a lack of publicly available data containing multi-level preference information; therefore, we collected such data ourselves to further evaluate the proposed approach. Specifically, we collected preference data for clothes using Prolific,\footnote{\url{https://www.prolific.com/}} a leading platform for recruiting participants to perform on-demand tasks for online research experiments and data collection. 

First, clothing items used for multi-level preference data collection were scraped from an online personal styling platform\footnote{\url{https://shoplook.io}} that allows users to discover outfit ideas or virtually create their own outfits. Items used in those outfits are typically linked to various third-party apparel-shopping websites. For the offline experiment, we selected 100 items, including 50 tops (for example, shirts or sweaters) and 50 bottoms (for example, jeans or pants) that appear among the most liked outfits on the platform. Because the majority of users on this personal styling website are women, the website focuses largely on women's fashion. Therefore, in this data collection, all selected 100 items represent women's clothing.

We then used Prolific to crowdsource preference ratings on all outfits (i.e., specific top-bottom combinations) and individual items (i.e., specific tops and bottoms). Specifically, we recruited 386 female participants\footnote{In this and following studies, informed content was obtained from all participants.} and each of them rated 42.1 outfits, 12.2 tops, and 11.8 bottoms on average. All ratings are on a discrete 1-to-5 scale (with 0.5 increment). With the collected data, we constructed the following data sources: utb-tensor of user-outfit (i.e., user-top-bottom) preferences ($386 \times 50 \times 50$) that has 16,254 elements (1.68\%) observed; ut-matrix of user-top preferences ($386 \times 50$) with 4,712 elements (24.4\%) observed; and ub-matrix of user-bottom preferences ($386 \times 50$) with 4,569 elements (23.7\%) observed. 

In addition, to measure the compatibility between items in an outfit and use it as an input of fashion expertise,
we built a database of objective fit ratings for all possible outfit combinations that can be made from the 50 tops and 50 bottoms. Specifically, we recruited another 300 subjects and asked each of them to evaluate the {\em degree of fit} (on a 5-point scale, irrespective of their own personal preference for the outfit) between the top and the bottom in 60 randomly generated outfits. 
The fit rating of each outfit was calculated as the average of its collected ratings—at least seven for each possible top-bottom combination. This data was used to construct tb-matrix of top-bottom compatibility indicators that are intended to measure the objective ``fit'' between two atomic items based on domain knowledge. 
The matrix size is $50 \times 50$, with 2,500 elements (100\%) observed.

The collected fashion data is used to evaluate the proposed method and other baselines as described in Section \ref{sec:sim3d}. For each of the four tensors (namely, utb-tensor, ut-matrix, ub-matrix, tb-matrix), we randomly split the data, allocating 75\% for training and 25\% for testing. 
The experimental evaluation is repeated 100 times with different random splits.


The experiment results using the standard RMSE metric are summarized in Table \ref{JIM_results}. (Corresponding  results using mean absolute error as evaluation metric as well as computational runtimes are provided in Table \ref{JIM_results-mae} of the Appendix.) From Table \ref{JIM_results} we can see that JIMA consistently improves upon the existing methods in nearly all cases. 
Furthermore, JIMA shows better performance compared to its ablated versions as well, indicating the importance of both joint modeling and interaction modeling capabilities. Specifically, the ablated JIMA outperforms its deep learning counterparts when relying solely on joint modeling, but shows no improvement when using only interactions. One possible reason is that the sample size of each tensor or matrix is not sufficiently large. Joint modeling alleviates the problem by learning across multi-level preference information, which in turn improves model training. It is also worth noting that, once joint modeling is in place, incorporating interactions further enhances JIMA’s performance. This aligns with our finding in Section \ref{sec:sim3d} with respect to the three data matrices, highlighting the value of both components in the proposed design.

Importantly, JIMA provides a solution for predicting user preferences on (and, hence, making recommendations of) outfits, tops, and bottoms jointly. In other words, while vast majority of existing methods are designed for single-level user preferences estimation (e.g., predicting item ratings based on ratings of some other items), JIMA is capable of leveraging {\em multi-level} preferences, e.g., predicting individual clothing item ratings (i.e., atomic item preferences) not just based on ratings of other individual clothing items but also on ratings of outfits (i.e., composite item preferences) and vice versa.  This can bring additional benefits in prediction accuracy, as demonstrated in all the aforementioned computational experiments.

\begin{table*}
	\begin{tiny}
		\caption{Outfit recommendation accuracy comparison between the proposed method (JIMA) and the competing methods on real fashion data using RMSE (with standard deviation in parentheses).}
		\vspace{-4mm}
		\begin{center}
			\begin{tabular}{cccccc}
				\hline
				\hline
& & \multicolumn{4}{c}{Data sources for measuring prediction performance} \\
\hline
				Model Type &Model Name &utb-tensor	 	  &ut-matrix	 	&ub-matrix	 &tb-matrix \\
				\hline
				Global Mean Imputation		&GMI & 1.126 (0.008) & 1.117 (0.015) & 1.177 (0.014) & 0.973 (0.020) \\
				\hline
				Tensor/Matrix &TF (CPD) &2.581 (0.254) & & & \\
				Factorization &MF & &1.006 (0.023) & & \\
				&MF & & &1.089 (0.031) & \\
				&MF & & & &0.799 (0.019) \\
				\hline
\vspace{1mm}
				Deep Learning &NTF (=NF$_{\mbox{utb}}$) &0.909 (0.013) & & & \\
\vspace{1mm}
				Benchmarks &NCF (=NF$_{\mbox{ut}}$) & &1.008 (0.030) & & \\
\vspace{1mm}
				&NCF (=NF$_{\mbox{ub}}$) & & &1.060 (0.022) & \\
\vspace{1mm}
				&NCF (=NF$_{\mbox{tb}}$) & & & & {\bf 0.658} (0.019) \\
				\hline
                \hline\\[-0.8em]
\vspace{1mm}
				JIMA &NF$^{\mbox{x}}_{\mbox{utb}\oplus\mbox{ut}\oplus\mbox{ub}\oplus\mbox{tb}}$ & {\bf 0.879} (0.013) & {\bf 0.974} (0.022) & {\bf 0.967} (0.026) & {\bf 0.658} (0.022) \\
				\hline
                \hline\\[-0.8em]
\vspace{1mm}
				Ablated JIMA &NF$^{\mbox{x}}_{\mbox{utb}}$ &0.911 (0.025) & & & \\
\vspace{1mm}
(interactions only)				&NF$^{\mbox{x}}_{\mbox{ut}}$ & &1.021 (0.027) & & \\
\vspace{1mm}				&NF$^{\mbox{x}}_{\mbox{ub}}$ & & &1.073 (0.028) & \\
\vspace{1mm}
&NF$^{\mbox{x}}_{\mbox{tb}}$ & & & &0.673 (0.167) \\
				\hline
\vspace{1mm}
Ablated JIMA &NF$_{\mbox{utb} \oplus\mbox{ut}}$ &0.896 (0.012) &1.001 (0.138) & & \\
\vspace{1mm}
(joint modeling only)	&NF$_{\mbox{utb} \oplus\mbox{ut}\oplus\mbox{ub}}$ &0.891 (0.012) &0.986 (0.022) &0.996 (0.028) & \\
\vspace{1mm}
				&NF$_{\mbox{utb} \oplus\mbox{ut}\oplus\mbox{ub}\oplus\mbox{tb}}$ &0.895 (0.067) &0.983 (0.021) &0.995 (0.029) &0.660 (0.029) \\
				\hline
				\hline
			\end{tabular}
		\end{center}
		\label{JIM_results}
	\end{tiny}
	\vspace{-1.5em}
\end{table*}

\section{Online Experiment} \label{sec:online}

There is a growing understanding in the recommender systems community that good performance on offline evaluation metrics does not always translate into good online performance \citep{garcin2014offline,beel2015comparison,rossetti2016contrasting,wang2023well}. To validate the performance advantages of JIMA in online settings, we ran additional studies on Prolific, where we asked human participants to rate outfits recommended by JIMA as well as by baseline techniques (including algorithmic or non-algorithmic methods) in a blinded test. 



The online experiments were designed to compare best {\em outfit} recommendations from several algorithmic and non-algorithmic baselines. For algorithmic baselines, the competing techniques include simple Linear Regression model (LR), Factorization Machine (FM) \citep{rendle2012factorization}, and Neural Tensor Factorization (NTF) \citep{wu2019neural}. Here LR and FM are general predictive modeling techniques that are commonly adopted across various recommender systems applications, and NTF is the most effective competing method for outfit recommendations based on our offline experiments in Sections \ref{sec:sim} and \ref{sec:offline}. 

We also benchmark JIMA against several non-algorithmic baselines. The first one OBJ recommends based on objective (i.e., non-personalized, user-independent) fit among clothing items within an outfit. This baseline represents a domain knowledge or expertise-based approach that is a natural option for composite-item recommendation when aesthetic item compatibility matters. Comparing to OBJ isolates the incremental value of integrating personal preference. The second baseline PRF aggregates subjective user preferences for an outfit, a preference driven approach similar to a population-consensus-based (i.e., average-rating-based) recommender commonly used in practice. Comparison against PRF can reveal the impact of fashion expertise (i.e., {\em degree of fit}) in driving recommendation performance. Finally, we include RDM (random recommendations) as an experimental control to gauge the improvement attributable to {\em data-driven} recommendations. 
All baselines, including algorithmic and non-algorithmic, were deployed online to generate outfit recommendations on the fly.

For this study, we used the same 100 women's clothing items as in our offline experiments, discussed in Section \ref{sec:offline}.  A total of 300 female participants were recruited from Prolific. Each of them was paid with a fixed fee upon completion of the study. No additional performance-based incentives were provided to participants. During the experiment, each participant was asked to complete the following four tasks.

\textbf{Task 1}: Each participant was presented with 24 images of different clothing items, i.e., 12 tops and 12 bottoms, randomly drawn from the pool of 100 clothing items, over 8 web pages. Participants were instructed to imagine they are shopping for clothes that they can wear and, for each clothing item, participants indicated how much they personally like it and would want to buy/wear it themselves by clicking on the accompanying radio buttons on the 5-point scale. An example web page for this task is displayed in Figure \ref{fig:task1} in the Appendix. The primary goal of this task is to collect participants' personal preferences for individual clothing items (assuming that each item is available in any size) to be used as inputs for clothing recommender systems.

\textbf{Task 2}: The participants were asked to go through 42 images of outfits over 14 web pages. Each outfit is a combination of a top and a bottom randomly selected from the same clothing item pool. Each participant was instructed to imagine they are shopping for outfits, that is, purchasing clothes that she can wear together. Given each outfit, the participants can indicate how much they personally like and would like to buy/wear the outfit themselves by clicking the corresponding radio button with the provided 5-point scale. An example of the web page for this task is shown in Figure \ref{fig:task2} in the Appendix. The goal of this task is to collect each participant's personal preferences for outfits that will be used as inputs for clothing recommender systems.

\begin{figure*}[htbp]
	\centering
	\captionsetup{width=\linewidth}
	\includegraphics[width=0.8\textwidth]{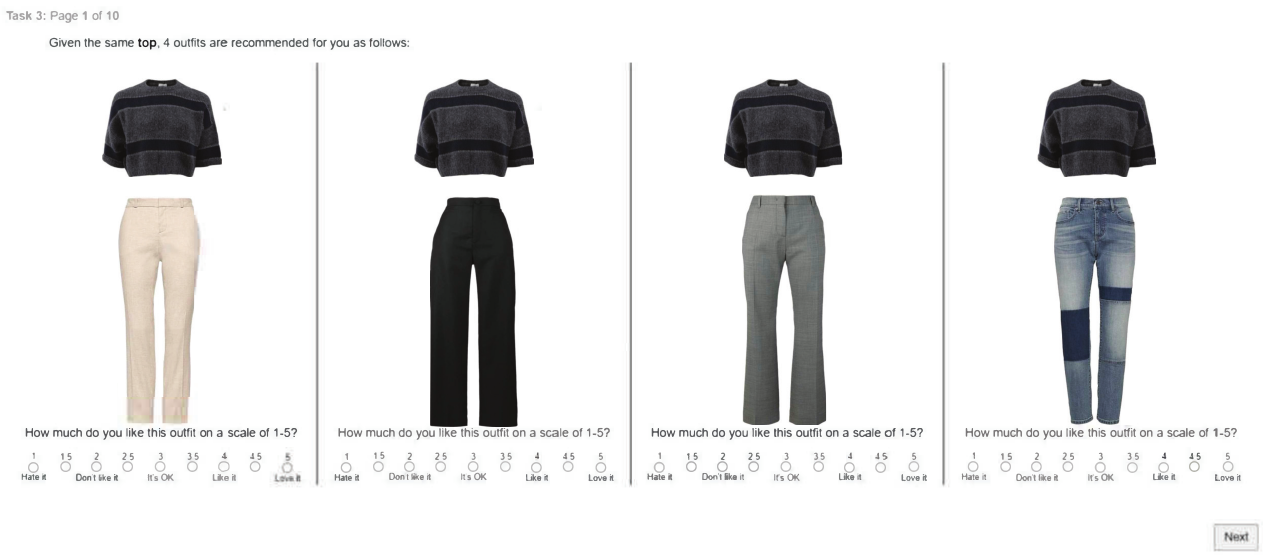}
	\centering
	\caption{\textbf{Example interface for recommendation rating using different techniques}}
	\label{fig:task3_s2}
	\vspace{-1.5em}
\end{figure*}

\textbf{Task 3}: The participants were asked to go through 10 web pages of outfit recommendations, each page containing four outfits. On each web page, one clothing item (bottom or top) was fixed (i.e., the same across all four outfits). Specifically, the 10 fixed items for each participant in this task were her 5 highest-rated tops and 5 highest-rated bottoms from Task 1. Given each highly rated top or bottom, outfit recommendations generated by four different techniques were displayed side by side, in random order. Given each recommendation, the participants were instructed to rate how much they personally like the recommendation and would like to buy/wear the outfit themselves by clicking the corresponding radio button with the provided 5-point scale. An example of the web page for this task is shown in Figure \ref{fig:task3_s2}. 

Each of the four recommendation methods was trained based on individual preference ratings reported in Tasks 1 and 2 combined with the data that we collected for offline experiments (Section \ref{sec:offline}). Each trained model was then used to predict ratings for all possible outfits containing the fixed item on each recommendation page and outfit with the highest predicted rating was presented for evaluation. When multiple methods provided the same recommendation, we displayed the next-best alternative (e.g., second- or third-ranked item) provided by JIMA as placeholders. Participants' preferences over placeholder outfits were not used in subsequent analysis.

As noted earlier, we compare JIMA to two sets of baselines, i.e., algorithmic (LR, FM, NTF) and non-algorithmic (OBJ, PRF, RDM). Of the 300 recruited participants, 150 were asked to evaluate recommendations from JIMA and the algorithmic baselines (i.e., Task 3a), while the remaining 150 evaluated JIMA alongside the non-algorithmic baselines (i.e., Task 3b). Participants were unaware of the underlying techniques, i.e., did not know which method generated each outfit. The goal was to identify which technique produced the most preferred outfit recommendations under a real-time evaluation.

\textbf{Task 4}: Participants were instructed to answer several survey questions on their demographics, background, and experiences with online clothes shopping, fashion, and online personalized recommendation services (See Column 1 of Table \ref{tab:demo}). Specifically, their experiences with fashion were measured by two constructs from prior work -- \textit{Fashion Opinion Leadership} \citep{goldsmith1991social,schrank1973correlates} and \textit{Clothing Interest} \citep{gurel1979clothing, creekmore1971methods} -- that are measured by aggregating participant responses from two sets of survey items, as presented in Table \ref{tab:fashion} of the Appendix. 

We applied a few data-quality and attention checks on the recruited participants. For instance, participants with incomplete responses, spent substantially less time on tasks, provided zero variance of preference ratings as well as inconsistent responses to fashion-related survey items\footnote{A reverse-coded item was added to the \textit{Fashion opinion leadership} scale, which also served as an attention check. Only participants whose responses aligned with the standard items after reverse scoring were kept.} were excluded from subsequent analysis. Descriptive statistics for the recruited participants after quality checks are displayed in Table \ref{tab:demo}. Column (2) reports 127 participants who evaluated recommendations from JIMA in comparison to the algorithmic baselines (i.e., Task 3a), while Column (3) reports 124 participants who evaluated JIMA against the non-algorithmic baselines (i.e., Task 3b). Comparisons of collected characteristics indicate participants are comparable across experimental settings.

\begin{table}[htbp]
	\centering\caption{\textbf{Descriptive Statistics on Participant Characteristics}}
	\scriptsize
		\begin{tabular}{llll}
			\hline
			&                           &\textbf{Task 3a} &\textbf{Task 3b}\\
			\hline
			Sample size                 &                           &127          &124\\[0.5em]
			Age: Mean (SD)              &                           &42.04 (13.10)&41.16 (14.17)\\[0.5em]
			Arts or Design              &Yes (\%)                   &13 (10.24\%) &9 (7.26\%) \\
			related occupation:         &No(\%)                     &114 (89.76\%)&115 (92.74\%)\\[0.5em]
			Online clothes              &Hate it(\%)                &4 (3.15\%)   &1 (0.81\%)\\
			shopping attitude:          &Don't like it(\%)          &12 (9.45\%)  &21(16.94\%)\\
			                            &Feel OK about it(\%)       &40 (31.50\%) &44(35.48\%)\\
			                            &Like it(\%)                &41 (32.28\%) &37(29.84\%) \\
			                            &Love it(\%)                &30 (23.62\%) &21(16.94\%) \\[0.5em]
			Online clothes              &Never(\%)                  &4  (3.15\%)  &2 (1.61\%) \\
			shopping frequency:         &$\leq 2$ time(s)/year(\%)  &25 (19.69\%) &36(29.03\%) \\
			                            &every 2 - 3 months(\%)     &50 (39.37\%) &52(41.94\%) \\
			                            &$\geq 1$ time(s)/month(\%) &40 (31.50\%) &29(23.39\%) \\
			                            &$\geq 1$ time(s)/week(\%)  &8  (6.30\%)  &5(4.03\%) \\[0.5em]
			\multicolumn{2}{l}{Below are constructs on a 1 - 7}     &             & \\
			\multicolumn{2}{l}{rating scale: Mean (SD)}             &             &\\ [0.5em]         
			\multicolumn{2}{l}{Fashion opinion leadership:}                     & 3.28 (1.72) &3.06 (1.53)\\
			\multicolumn{2}{l}{Clothing interest:}                              & 3.64 (1.62) &3.28 (1.41)\\
			\multicolumn{2}{l}{Experience with online recommendation:}          & 5.44 (1.45) &4.85 (1.77)\\
			\multicolumn{2}{l}{Perception of online recommendation usefulness:} & 5.05 (1.40) &5.04 (1.40)\\
			\hline
		\end{tabular}
        \begin{minipage}{1\linewidth}
        \scriptsize
         \textit{Notes:} Participants in Task 3a were asked to rate outfits recommended by 
         JIMA and its algorithmic baselines (i.e., NTF, FM, LR).\\ 
         Participants in Task 3b were asked to rate outfits recommended by 
         JIMA and its non-algorithmic baselines (i.e., OBJ, PRF, RDM).
       \end{minipage}
	\label{tab:demo}
\end{table}


Our key analysis compares the outfit ratings each participant provided in Task 3. Table \ref{tab:s2_recrating} reports the mean ratings for outfits recommended by the four methods: JIMA, NTF, FM, and LR. On average, participants rated JIMA's recommendations higher than all algorithmic baselines, and the differences between JIMA and each of the baselines are substantial and highly statistically significant. In short, JIMA recommends outfits that are better aligned with participants' individual preferences.
\begin{table}[htbp]
	\centering\caption{\textbf{Recommendation Performance Comparison with Algorithmic Baselines}}
	\resizebox{0.47\textwidth}{!}{%
		\begin{tabular}{lccc}
			\hline
			\textbf{Technique}    &\textbf{\#Observations} &\textbf{Mean (SD)}  &\textbf{t-statistic}\\
			\hline
			JIMA          &1270           &{\bf 3.292 (0.030)} &\\ 
			NTF           &1270           &3.198 (0.030) &3.567***\\ 
			FM            &1270           &2.859 (0.032) &12.132***\\ 
			LR            &1270           &2.722 (0.031) &15.873***\\ 
			\hline
			\multicolumn{4}{l}{\footnotesize{$t$-test: mean(JIMA) $>$ mean(Baseline)}}\\
            \multicolumn{4}{l}{\footnotesize{*$p$ $\leq$ 0.05, **$p$ $\leq$ 0.01, ***$p$ $\leq$ 0.001}}
	\end{tabular}
} 
\label{tab:s2_recrating}
\end{table}

The performance comparison between JIMA and the non-algorithmic baselines is displayed in Table \ref{tab:nonalg}. This set of results shows that JIMA produces significantly more satisfying outfit recommendations than (i) the pure domain-expertise-based method that considers only compatibility between individual items, (ii) the aggregate preference-based method that relies solely on participants' subjective outfit preferences, and (iii) random recommendations. Taken together with the previous results against the algorithmic baselines (Table \ref{tab:s2_recrating}), these findings further highlight the practical value of multi-level preference information and the advantages of joint interaction modeling for composite-item recommendation.

\begin{table}[htbp]
	\centering\caption{\textbf{Recommendation Performance Comparison with Non-Algorithmic Baselines}}
	\resizebox{0.47\textwidth}{!}{%
		\begin{tabular}{lccc}
			\hline
			\textbf{Technique}    &\textbf{\#Observations} &\textbf{Mean (SD)}  &\textbf{t-statistic}\\
			\hline
			JIMA          &1240           &{\bf 3.212 (1.059)}\\ 
			OBJ           &1240           &2.990 (1.127) &6.807***\\ 
			PRF           &1240           &2.800 (1.140) &11.777***\\ 
			RDM           &1240           &2.777 (1.108) &12.048***\\ 
			\hline
			\multicolumn{4}{l}{\footnotesize{$t$-test: mean(JIMA) $>$ mean(Baseline)}} \\
            \multicolumn{4}{l}{\footnotesize{*$p$ $\leq$ 0.05, **$p$ $\leq$ 0.01, ***$p$ $\leq$ 0.001.}}
	\end{tabular}
} 
\label{tab:nonalg}
	\vspace{-1em}
\end{table}

\section{Discussion and Conclusions} \label{sec:discussion}


We propose a novel deep-learning-based recommendation approach that provides effective recommendations by jointly modeling multi-level user preferences and leveraging latent interactions to address the challenging compatibility issue, where a user may like atomic items (e.g., a top and a bottom) but not the composite item (e.g., an outfit). The proposed method is a systematic approach that is able to take advantage of multi-level preference information that can simultaneously include and model not only subjective, user-specific preferences for atomic and composite items but also objective (e.g., domain-expertise-based) information on which items fit together better. The ability to learn a unified underlying model also allows making recommendations at different item granularities. 


The performance advantages of the proposed approach are demonstrated using comprehensive computational simulations, offline experiments with real-world data, and online experiments with real users in the context of a fashion recommender system. In simulation studies, the proposed method demonstrates excellent performance across a variety of complex settings (multi-level preferences that include two-item and three-item outfits, different data sizes, presence of objective item fit information, etc.), substantially outperforming the baselines. In the offline experiment with real data, the proposed method continues to outperform existing benchmarks for both outfits and individual item recommendations. Furthermore, the online experiment evaluates the proposed method and the baseline directly with human participants. Participants consistently assigned higher ratings to outfits recommended by the proposed method than those provided by all the competing approaches (both algorithmic and non-algorithmic). In summary, across both offline and online evaluation settings, the proposed method demonstrates consistent performance gains for composite-item recommendation.

Importantly, although we use fashion recommendation as the illustrative setting, the approach can be naturally generalizable to other applications with complex, multi-level preference information.  One example is {\em context-aware}  recommender systems  \citep{adomavicius2011context, Adomavicius2022}, where the proposed approach could model the composite effects of multiple contextual factors on user preferences. More specifically, one core assumption in context-aware recommender systems is that user preferences for items vary by context. Thus, user preference modeling is typically enhanced by incorporating {\em contextual} factors, in addition to user and item features, that impacts user preferences for
items.  For example, users $u$ generally have preferences for different songs $s$; such preferences could be represented by a $us$-matrix. Furthermore, preferences may also be context-specific: users $u$'s evaluation of songs $s$ can depend on their mood $m$ (happy, sad, reflective, etc.), represented by a $usm$-tensor, or on their activity $a$ (working, exercising, resting, etc.), represented by a $usa$-tensor. In more complex scenarios, users $u$ may express even more nuanced preferences for songs $s$ with both mood $m$ and activity $a$, producing a $usma$-tensor. A music streaming service may have historical preference data at various levels of granularity, and JIMA can naturally learn a unifying model from these multi-level preferences and provide preference predictions and recommendations at any level.



Therefore, future studies can investigate the applicability and performance of the proposed approach in more complex settings, e.g., fashion contexts where outfits comprise more than two or three atomic items, or in completely different application domains. For the latter, one natural candidate is context-aware recommender systems, as mentioned earlier. Further directions include more extensive simulation studies and applications to additional real-world datasets.



\bibliographystyle{IEEEtran}
\bibliography{joint_model}

@inproceedings{beel2015comparison,
	author = {Beel, Joeran and Langer, Stefan},
	booktitle = {International conference on theory and practice of digital libraries},
	organization = {Springer},
	pages = {153--168},
	title = {A comparison of offline evaluations, online evaluations, and user studies in the context of research-paper recommender systems},
	year = {2015}}

@inproceedings{garcin2014offline,
	author = {Garcin, Florent and Faltings, Boi and Donatsch, Olivier and Alazzawi, Ayar and Bruttin, Christophe and Huber, Amr},
	booktitle = {Proceedings of the 8th ACM Conference on Recommender systems},
	pages = {169--176},
	title = {Offline and online evaluation of news recommender systems at swissinfo.ch},
	year = {2014}}

@inproceedings{rossetti2016contrasting,
	author = {Rossetti, Marco and Stella, Fabio and Zanker, Markus},
	booktitle = {Proceedings of the 10th ACM conference on recommender systems},
	pages = {31--34},
	title = {Contrasting offline and online results when evaluating recommendation algorithms},
	year = {2016}}

@inproceedings{wang2023well,
	author = {Wang, Xiaojie and Gao, Ruoyuan and Jain, Anoop and Edge, Graham and Ahuja, Sachin},
	booktitle = {Proc. of the 46th Intl. ACM SIGIR Conf. on Research and Development in Information Retrieval},
	pages = {3415--3420},
	title = {How well do offline metrics predict online performance of product ranking models?},
	year = {2023}}

@inproceedings{chen2019deep,
	author = {Chen, Zhengyu and Gai, Sibo and Wang, Donglin},
	booktitle = {2019 IEEE Intl. Conf. on Big Data},
	organization = {IEEE},
	pages = {1046--1051},
	title = {Deep tensor factorization for multi-criteria recommender systems},
	year = {2019}}

@inproceedings{lu2021personalized,
	author = {Lu, Zhi and Hu, Yang and Chen, Yan and Zeng, Bing},
	booktitle = {Proceedings of the IEEE/CVF conference on computer vision and pattern recognition},
	pages = {12722--12731},
	title = {Personalized outfit recommendation with learnable anchors},
	year = {2021}}

@article{frolov2017tensor,
	author = {Frolov, Evgeny and Oseledets, Ivan},
	journal = {Wiley Interdisciplinary Reviews: Data Mining and Knowledge Discovery},
	number = {3},
	pages = {e1201},
	publisher = {Wiley Online Library},
	title = {Tensor methods and recommender systems},
	volume = {7},
	year = {2017}}

@article{verbert2012context,
	author = {Verbert, Katrien and Manouselis, Nikos and Ochoa, Xavier and Wolpers, Martin and Drachsler, Hendrik and Bosnic, Ivana and Duval, Erik},
	journal = {IEEE transactions on learning technologies},
	number = {4},
	pages = {318--335},
	publisher = {IEEE},
	title = {Context-aware recommender systems for learning: a survey and future challenges},
	volume = {5},
	year = {2012}}

@article{bi2022improving,
	author = {Bi, Xuan and Adomavicius, Gediminas and Li, William and Qu, Annie},
	journal = {INFORMS Journal on Computing},
	number = {3},
	pages = {1644--1660},
	publisher = {INFORMS},
	title = {Improving sales forecasting accuracy: a tensor factorization approach with demand awareness},
	volume = {34},
	year = {2022}}

@article{zhang2021dynamic,
	author = {Zhang, Yanqing and Bi, Xuan and Tang, Niansheng and Qu, Annie},
	journal = {Journal of machine learning research},
	number = {65},
	pages = {1--35},
	title = {Dynamic tensor recommender systems},
	volume = {22},
	year = {2021}}

@article{bi2021tensors,
	author = {Bi, Xuan and Tang, Xiwei and Yuan, Yubai and Zhang, Yanqing and Qu, Annie},
	journal = {Annual review of statistics and its application},
	pages = {345--368},
	publisher = {Annual Reviews},
	title = {Tensors in statistics},
	volume = {8},
	year = {2021}}

@inproceedings{sun2019bert4rec,
	author = {Sun, Fei and Liu, Jun and Wu, Jian and Pei, Changhua and Lin, Xiao and Ou, Wenwu and Jiang, Peng},
	booktitle = {Proc. of the 28th ACM Intl. Conf. on information and knowledge management},
	pages = {1441--1450},
	title = {BERT4Rec: Sequential recommendation with bidirectional encoder representations from transformer},
	year = {2019}}

@inproceedings{kang2018self,
	author = {Kang, Wang-Cheng and McAuley, Julian},
	booktitle = {2018 IEEE international conference on data mining (ICDM)},
	organization = {IEEE},
	pages = {197--206},
	title = {Self-attentive sequential recommendation},
	year = {2018}}

@article{wu2022graph,
	author = {Wu, Shiwen and Sun, Fei and Zhang, Wentao and Xie, Xu and Cui, Bin},
	journal = {ACM Computing Surveys},
	number = {5},
	pages = {1--37},
	publisher = {ACM New York, NY},
	title = {Graph neural networks in recommender systems: a survey},
	volume = {55},
	year = {2022}}

@article{gao2023survey,
	author = {Gao, Chen and Zheng, Yu and Li, Nian and Li, Yinfeng and Qin, Yingrong and Piao, Jinghua and Quan, Yuhan and Chang, Jianxin and Jin, Depeng and He, Xiangnan and others},
	journal = {ACM Transactions on Recommender Systems},
	number = {1},
	pages = {1--51},
	publisher = {ACM New York, NY, USA},
	title = {A survey of graph neural networks for recommender systems: Challenges, methods, and directions},
	volume = {1},
	year = {2023}}

@inproceedings{hu2015collaborative,
	author = {Hu, Yang and Yi, Xi and Davis, Larry S},
	booktitle = {Proceedings of the 23rd ACM international conference on Multimedia},
	pages = {129--138},
	title = {Collaborative fashion recommendation: A functional tensor factorization approach},
	year = {2015}}

@inproceedings{liu2012hi,
	author = {Liu, Si and Feng, Jiashi and Song, Zheng and Zhang, Tianzhu and Lu, Hanqing and Xu, Changsheng and Yan, Shuicheng},
	booktitle = {Proceedings of the 20th ACM international conference on Multimedia},
	pages = {619--628},
	title = {Hi, magic closet, tell me what to wear!},
	year = {2012}}

@inproceedings{li2020hierarchical,
	author = {Li, Xingchen and Wang, Xiang and He, Xiangnan and Chen, Long and Xiao, Jun and Chua, Tat-Seng},
	booktitle = {Proc. of the 43rd Intl. ACM SIGIR Conf. on Research and Development in Information Retrieval},
	pages = {159--168},
	title = {Hierarchical fashion graph network for personalized outfit recommendation},
	year = {2020}}

@inproceedings{wu2021self,
	author = {Wu, Jiancan and Wang, Xiang and Feng, Fuli and He, Xiangnan and Chen, Liang and Lian, Jianxun and Xie, Xing},
	booktitle = {Proceedings of the 44th international ACM SIGIR conference on research and development in information retrieval},
	pages = {726--735},
	title = {Self-supervised graph learning for recommendation},
	year = {2021}}

@article{li2017mining,
	author = {Li, Yuncheng and Cao, Liangliang and Zhu, Jiang and Luo, Jiebo},
	journal = {IEEE Transactions on Multimedia},
	number = {8},
	pages = {1946--1955},
	publisher = {IEEE},
	title = {Mining fashion outfit composition using an end-to-end deep learning approach on set data},
	volume = {19},
	year = {2017}}

@inproceedings{mcauley2015image,
	author = {McAuley, Julian and Targett, Christopher and Shi, Qinfeng and Van Den Hengel, Anton},
	booktitle = {Proceedings of the 38th international ACM SIGIR conference on research and development in information retrieval},
	pages = {43--52},
	title = {Image-based recommendations on styles and substitutes},
	year = {2015}}

@inproceedings{veit2015learning,
	author = {Veit, Andreas and Kovacs, Balazs and Bell, Sean and McAuley, Julian and Bala, Kavita and Belongie, Serge},
	booktitle = {Proceedings of the IEEE international conference on computer vision},
	pages = {4642--4650},
	title = {Learning visual clothing style with heterogeneous dyadic co-occurrences},
	year = {2015}}

@inproceedings{vasileva2018learning,
	author = {Vasileva, Mariya I and Plummer, Bryan A and Dusad, Krishna and Rajpal, Shreya and Kumar, Ranjitha and Forsyth, David},
	booktitle = {Proceedings of the European conference on computer vision (ECCV)},
	pages = {390--405},
	title = {Learning type-aware embeddings for fashion compatibility},
	year = {2018}}

@inproceedings{chen2019pog,
	author = {Chen, Wen and Huang, Pipei and Xu, Jiaming and Guo, Xin and Guo, Cheng and Sun, Fei and Li, Chao and Pfadler, Andreas and Zhao, Huan and Zhao, Binqiang},
	booktitle = {Proceedings of the 25th ACM SIGKDD international conference on knowledge discovery \& data mining},
	pages = {2662--2670},
	title = {POG: personalized outfit generation for fashion recommendation at Alibaba iFashion},
	year = {2019}}

@inproceedings{han2017learning,
	author = {Han, Xintong and Wu, Zuxuan and Jiang, Yu-Gang and Davis, Larry S},
	booktitle = {Proceedings of the 25th ACM international conference on Multimedia},
	pages = {1078--1086},
	title = {Learning fashion compatibility with bidirectional lstms},
	year = {2017}}

@article{kuhn2019supporting,
	author = {Kuhn, Tobias and Bourke, Steven and Brinkmann, Levin and Buchwald, Tobias and Digan, Conor and Hache, Hendrik and Jaeger, Sebastian and Lehmann, Patrick and Maier, Oskar and Matting, Stefan and others},
	journal = {arXiv preprint arXiv:1908.09493},
	title = {Supporting stylists by recommending fashion style},
	year = {2019}}

@inproceedings{zhao2017deep,
	author = {Zhao, Kui and Hu, Xia and Bu, Jiajun and Wang, Can},
	booktitle = {Workshops at the 31st AAAI Conf. on Artificial Intelligence},
	title = {Deep style match for complementary recommendation},
	year = {2017}}

@article{wu2019session,
	author = {Wu, Jui-Chieh and Rodr{\'\i}guez, Jos{\'e} Antonio S{\'a}nchez and Pamp{\'\i}n, Humberto Jes{\'u}s Corona},
	journal = {arXiv preprint arXiv:1908.08327},
	title = {Session-based complementary fashion recommendations},
	year = {2019}}

@inproceedings{yamaguchi2012parsing,
	author = {Yamaguchi, Kota and Kiapour, M Hadi and Ortiz, Luis E and Berg, Tamara L},
	booktitle = {2012 IEEE Conference on Computer vision and pattern recognition},
	organization = {IEEE},
	pages = {3570--3577},
	title = {Parsing clothing in fashion photographs},
	year = {2012}}

@article{liang2016clothes,
	author = {Liang, Xiaodan and Lin, Liang and Yang, Wei and Luo, Ping and Huang, Junshi and Yan, Shuicheng},
	journal = {IEEE Transactions on Multimedia},
	number = {6},
	pages = {1175--1186},
	publisher = {IEEE},
	title = {Clothes co-parsing via joint image segmentation and labeling with application to clothing retrieval},
	volume = {18},
	year = {2016}}

@inproceedings{zhang2018quality,
	author = {Zhang, Yin and Lu, Haokai and Niu, Wei and Caverlee, James},
	booktitle = {Proceedings of the 12th ACM conference on recommender systems},
	pages = {77--85},
	title = {Quality-aware neural complementary item recommendation},
	year = {2018}}

@article{lasserre2018studio2shop,
	author = {Lasserre, Julia and Rasch, Katharina and Vollgraf, Roland},
	journal = {arXiv preprint arXiv:1807.00556},
	title = {Studio2shop: from studio photo shoots to fashion articles},
	year = {2018}}

@article{bracher2016fashion,
	author = {Bracher, Christian and Heinz, Sebastian and Vollgraf, Roland},
	journal = {arXiv preprint arXiv:1609.02489},
	title = {Fashion DNA: merging content and sales data for recommendation and article mapping},
	year = {2016}}

@inproceedings{kang2017visually,
	author = {Kang, Wang-Cheng and Fang, Chen and Wang, Zhaowen and McAuley, Julian},
	booktitle = {2017 IEEE Intl. Conf. on Data Mining (ICDM)},
	organization = {IEEE},
	pages = {207--216},
	title = {Visually-aware fashion recommendation and design with generative image models},
	year = {2017}}

@inproceedings{kato2019gans,
	author = {Kato, Natsumi and Osone, Hiroyuki and Oomori, Kotaro and Ooi, Chun Wei and Ochiai, Yoichi},
	booktitle = {Proc. of the 10th Augmented Human International Conf.},
	pages = {1--7},
	title = {{GAN}s-based clothes design: Pattern maker is all you need to design clothing},
	year = {2019}}

@article{guo2016deep,
	author = {Guo, Yanming and Liu, Yu and Oerlemans, Ard and Lao, Songyang and Wu, Song and Lew, Michael S},
	journal = {Neurocomputing},
	pages = {27--48},
	publisher = {Elsevier},
	title = {Deep learning for visual understanding: A review},
	volume = {187},
	year = {2016}}

@inproceedings{deng2020personalized,
	author = {Deng, Qilin and Wang, Kai and Zhao, Minghao and Zou, Zhene and Wu, Runze and Tao, Jianrong and Fan, Changjie and Chen, Liang},
	booktitle = {Proceedings of the 29th ACM International Conference on Information \& Knowledge Management},
	pages = {2381--2388},
	title = {Personalized bundle recommendation in online games},
	year = {2020}}

@inproceedings{zhu2014bundle,
	author = {Zhu, Tao and Harrington, Patrick and Li, Junjun and Tang, Lei},
	booktitle = {Proc. of the 37th Intl. ACM SIGIR Conf. on Research \& development in information retrieval},
	pages = {657--666},
	title = {Bundle recommendation in ecommerce},
	year = {2014}}

@inproceedings{bai2019personalized,
	author = {Bai, Jinze and Zhou, Chang and Song, Junshuai and Qu, Xiaoru and An, Weiting and Li, Zhao and Gao, Jun},
	booktitle = {The World Wide Web Conference},
	pages = {60--71},
	title = {Personalized bundle list recommendation},
	year = {2019}}

@article{guo2017deepfm,
	author = {Guo, Huifeng and Tang, Ruiming and Ye, Yunming and Li, Zhenguo and He, Xiuqiang},
	journal = {arXiv preprint arXiv:1703.04247},
	title = {DeepFM: a factorization-machine based neural network for CTR prediction},
	year = {2017}}

@article{gurel1979clothing,
	author = {Gurel, Lois M and Gurel, Lee},
	journal = {Home Economics Research Journal},
	number = {5},
	pages = {274--282},
	publisher = {Wiley Online Library},
	title = {Clothing interest: Conceptualization and measurement},
	volume = {7},
	year = {1979}}

@article{goldsmith1991social,
	author = {Goldsmith, Ronald E and Heitmeyer, Jeanne R and Freiden, Jon B},
	journal = {Clothing and Textiles Research Journal},
	number = {1},
	pages = {37--45},
	publisher = {Sage Publications Sage CA: Thousand Oaks, CA},
	title = {Social values and fashion leadership},
	volume = {10},
	year = {1991}}

@article{schrank1973correlates,
	author = {Schrank, Holly L and Lois Gilmore, D},
	journal = {Sociological Quarterly},
	number = {4},
	pages = {534--543},
	publisher = {Wiley Online Library},
	title = {Correlates of fashion leadership: Implications for fashion process theory},
	volume = {14},
	year = {1973}}

@article{creekmore1971methods,
	author = {Creekmore, AM},
	journal = {Michigan State University},
	title = {Methods of measuring clothing variables (Michigan Agricultural Experiment Station Project No. 783)},
	year = {1971}}

@inproceedings{yu2019complementary,
	author = {Yu, Hang and Litchfield, Lester and Kernreiter, Thomas and Jolly, Seamus and Hempstalk, Kathryn},
	booktitle = {2019 International Conference on High Performance Big Data and Intelligent Systems (HPBD\&IS)},
	organization = {IEEE},
	pages = {73--78},
	title = {Complementary recommendations: A brief survey},
	year = {2019}}

@inproceedings{wu2019neural,
	author = {Wu, Xian and Shi, Baoxu and Dong, Yuxiao and Huang, Chao and Chawla, Nitesh V},
	booktitle = {Proc. of the Twelfth ACM Intl. Conf. on Web Search and Data Mining},
	pages = {537--545},
	title = {Neural tensor factorization for temporal interaction learning},
	year = {2019}}

@article{kingma2014adam,
	author = {Kingma, Diederik P and Ba, Jimmy},
	journal = {arXiv preprint arXiv:1412.6980},
	title = {Adam: A method for stochastic optimization},
	year = {2014}}

@inproceedings{He2017,
	author = {He, Xiangnan and Liao, Lizi and Zhang, Hanwang and Nie, Liqiang and Hu, Xia and Chua, Tat-Seng},
	booktitle = {Proc. 26th Int. Conf. world wide web},
	pages = {173--182},
	title = {{Neural collaborative filtering}},
	year = {2017}}

@article{bi2018multilayer,
	author = {Bi, Xuan and Qu, Annie and Shen, Xiaotong and others},
	journal = {The Annals of Statistics},
	number = {6B},
	pages = {3308--3333},
	publisher = {Institute of Mathematical Statistics},
	title = {Multilayer tensor factorization with applications to recommender systems},
	volume = {46},
	year = {2018}}

@article{bi2017group,
	author = {Bi, Xuan and Qu, Annie and Wang, Junhui and Shen, Xiaotong},
	journal = {Journal of the American Statistical Association},
	number = {519},
	pages = {1344--1353},
	publisher = {Taylor \& Francis},
	title = {A group-specific recommender system},
	volume = {112},
	year = {2017}}

@book{ricci2022recommender,
	author = {Ricci, Francesco and Rokach, Lior and Shapira, Bracha},
	edition = {3rd},
	isbn = {1489976361, 9781489976369},
	publisher = {Springer},
	title = {Recommender Systems Handbook},
	year = {2022}}

@article{adomavicius2005toward,
	author = {Adomavicius, Gediminas and Tuzhilin, Alexander},
	journal = {IEEE Transactions on Knowledge and Data Engineering},
	number = {6},
	pages = {734--749},
	publisher = {IEEE},
	title = {Toward the next generation of recommender systems: A survey of the state-of-the-art and possible extensions},
	volume = {17},
	year = {2005}}

@incollection{adomavicius2011context,
	author = {Adomavicius, Gediminas and Tuzhilin, Alexander},
	booktitle = {Recommender Systems Handbook},
	pages = {217--253},
	publisher = {Springer},
	title = {Context-aware recommender systems},
	year = {2011}}

@inproceedings{agarwal2009regression,
	author = {Agarwal, Deepak and Chen, Bee-Chung},
	booktitle = {Proceedings of the 15th ACM SIGKDD International Conference on Knowledge Discovery and Data Mining},
	organization = {ACM},
	pages = {19--28},
	title = {Regression-based latent factor models},
	year = {2009}}

@inproceedings{bell2007scalable,
	author = {Bell, Robert M and Koren, Yehuda},
	booktitle = {Proceedings of the 2007 7th IEEE International Conference on Data Mining},
	organization = {IEEE},
	pages = {43--52},
	title = {Scalable collaborative filtering with jointly derived neighborhood interpolation weights},
	year = {2007}}

@article{carroll1970analysis,
	author = {Carroll, J Douglas and Chang, Jih-Jie},
	journal = {Psychometrika},
	number = {3},
	pages = {283--319},
	publisher = {Springer},
	title = {Analysis of individual differences in multidimensional scaling via an {N}-way generalization of ``{E}ckart-{Y}oung'' decomposition},
	volume = {35},
	year = {1970}}

@article{funk2006netflix,
	author = {Funk, Simon},
	journal = {\url{http://sifter.org/~simon/journal/20061211.html}},
	title = {Netflix update: Try this at home},
	year = {2006}}

@inproceedings{lang1995newsweeder,
	author = {Lang, Ken},
	booktitle = {Proceedings of the 12th International Conference on Machine Learning},
	pages = {331--339},
	title = {Newsweeder: Learning to filter netnews},
	year = {1995}}

@inproceedings{mooney2000content,
	author = {Mooney, Raymond J and Roy, Loriene},
	booktitle = {Proceedings of the 5th ACM Conference on Digital Libraries},
	organization = {ACM},
	pages = {195--204},
	title = {Content-based book recommending using learning for text categorization},
	year = {2000}}

@article{nguyen2013content,
	author = {Nguyen, Jennifer and Zhu, Mu},
	journal = {Statistical Analysis and Data Mining: The ASA Data Science Journal},
	number = {4},
	pages = {286--301},
	publisher = {Wiley Online Library},
	title = {Content-boosted matrix factorization techniques for recommender systems},
	volume = {6},
	year = {2013}}

@article{rendle2012factorization,
	author = {Rendle, Steffen},
	journal = {ACM Transactions on Intelligent Systems and Technology (TIST)},
	number = {3},
	pages = {57},
	publisher = {ACM},
	title = {Factorization machines with {libFM}},
	volume = {3},
	year = {2012}}

@inproceedings{salakhutdinov2007restricted,
	author = {Salakhutdinov, Ruslan and Mnih, Andriy and Hinton, Geoffrey},
	booktitle = {Proceedings of the 24th International Conference on Machine Learning},
	organization = {ACM},
	pages = {791--798},
	title = {Restricted {B}oltzmann machines for collaborative filtering},
	year = {2007}}

@incollection{Adomavicius2022,
	address = {New York, NY},
	author = {Adomavicius, Gediminas and Bauman, Konstantin and Tuzhilin, Alexander and Unger, Moshe},
	booktitle = {Recommender Systems Handbook},
	editor = {Ricci, Francesco and Rokach, Lior and Shapira, Bracha},
	pages = {211--250},
	publisher = {Springer US},
	title = {Context-Aware Recommender Systems: From Foundations to Recent Developments},
	year = {2022}}

\clearpage

\onecolumn

\section*{Appendix: Recommending Composite Items Using Multi-Level Preference Information:\\ A Joint Interaction Modeling Approach}

\setcounter{table}{0}
\renewcommand{\thetable}{A\arabic{table}}
\setcounter{figure}{0}
\renewcommand{\thefigure}{A\arabic{figure}}
\setcounter{page}{1}

\subsection*{Additional Results and Evaluation of Simulation Study 1}

First, we provide additional simulation results for Section \ref{sec:sim3d}. The results are measured in root mean square errors (RMSE) in Tables \ref{Sim50-25-75}, \ref{Sim75-100-80}, and \ref{Sim150-100-80}, corresponding to $(N,T,B)$ being $(50,25,75)$, $(75,100,80)$, and $(150,100,80)$, respectively.
Next, we provide all results in mean absolute errors (MAE) as well as the computational time used for each method. The results are provided in Tables \ref{Sim100-100-100-mae}, \ref{Sim50-25-75-mae}, \ref{Sim75-100-80-mae}, and \ref{Sim150-100-80-mae}, corresponding to $(N,T,B)$ being $(100,100,100)$ $(50,25,75)$, $(75,100,80)$, and $(150,100,80)$, respectively. It can be seen that the findings are consistent with the results in the main text. All computations were done on a MacBook Pro with Apple M2 Max CPU.

\begin{table}[H]
\begin{tiny}
\caption{Simulated outfit recommendation accuracy comparison between the proposed methods (JIMA) and the competing methods with $N=50$, $T=25$, and $B=75$ using RMSE (with standard deviation in parentheses). The best performance is highlighted in bold.}
\vspace{-4mm}
\begin{center}
\begin{tabular}{cccccc}
\hline
\hline
& & \multicolumn{4}{c}{Data sources for measuring prediction performance} \\
\hline
Model Type & Model Name & utb-tensor & ut-matrix & ub-matrix & tb-matrix \\
\hline
Global Mean Imputation & GMI & 5.558 (0.327) & 2.270 (0.158) & 2.251 (0.139) & 2.283 (0.161) \\
\hline
Tensor/Matrix Factorization & TF (CPD) & 3.236 (0.209) &  &  &  \\
& MF &  & 2.245 (0.150) &  &  \\
& MF &  &  & 2.213 (0.142) &  \\
& MF &  &  &  & 2.254 (0.162) \\
\hline
Deep Learning Benchmarks & NTF (=NF$_{\mbox{utb}}$) & 0.894 (0.274) &  &  &  \\
& NCF (=NF$_{\mbox{ut}}$) &  & 2.533 (0.145) &  &  \\
& NCF (=NF$_{\mbox{ub}}$) &  &  & 2.473 (0.170) &  \\
& NCF (=NF$_{\mbox{tb}}$) &  &  &  & 2.598 (0.226) \\
\hline
\hline
JIMA & NF$^{\mbox{x}}_{\mbox{utb}\oplus\mbox{ut}\oplus\mbox{ub}\oplus\mbox{tb}}$ & 0.385 (0.059) & \textbf{0.555 (0.158)} & \textbf{0.416 (0.294)} & \textbf{0.456 (0.141)} \\
\hline
\hline
\vspace{1mm}
Ablated JIMA (interactions only) & NF$^{\mbox{x}}_{\mbox{utb}}$ & \textbf{0.343 (0.068)} &  &  &  \\
\vspace{1mm}
& NF$^{\mbox{x}}_{\mbox{ut}}$ &  & 2.524 (0.170) &  &  \\
\vspace{1mm}
& NF$^{\mbox{x}}_{\mbox{ub}}$ &  &  & 2.503 (0.163) &  \\
& NF$^{\mbox{x}}_{\mbox{tb}}$ &  &  &  & 2.601 (0.215) \\
\hline
Ablated JIMA (joint modeling only) & NF$_{\mbox{utb}\oplus\mbox{ut}}$ & 0.869 (0.186) & 1.162 (0.207) &  &  \\
& NF$_{\mbox{utb}\oplus\mbox{ut}\oplus\mbox{ub}}$ & 1.000 (0.343) & 1.224 (0.248) & 0.864 (0.249) &  \\
& NF$_{\mbox{utb}\oplus\mbox{ut}\oplus\mbox{ub}\oplus\mbox{tb}}$ & 1.032 (0.324) & 1.218 (0.184) & 0.912 (0.265) & 1.035 (0.198) \\
\hline
\hline
\end{tabular}
\end{center}
\label{Sim50-25-75}
\end{tiny}
\end{table}


\begin{table}[H]
\begin{tiny}
\caption{Simulated outfit recommendation accuracy comparison between the proposed methods (JIMA) and the competing methods with $N=75$, $T=100$, and $B=80$ using RMSE (with standard deviation in parentheses). The best performance is highlighted in bold.}
\vspace{-4mm}
\begin{center}
\begin{tabular}{cccccc}
\hline\hline
& & \multicolumn{4}{c}{Data sources for measuring prediction performance} \\
\hline
Model Type & Model Name & utb-tensor & ut-matrix & ub-matrix & tb-matrix \\
\hline
Global Mean Imputation & GMI & 5.490 (0.243) & 2.258 (0.123) & 2.238 (0.114) & 2.238 (0.111) \\
\hline
Tensor/Matrix Factorization & TF (CPD) & 3.126 (0.145) &  &  &  \\
& MF &  & 2.148 (0.129) &  &  \\
& MF &  &  & 2.178 (0.119) &  \\
& MF &  &  &  & 2.129 (0.123) \\
\hline
Deep Learning Benchmarks & NTF (=NF$_{\mbox{utb}}$) & 0.468 (0.031) &  &  &  \\
& NCF (=NF$_{\mbox{ut}}$) &  & 2.367 (0.172) &  &  \\
& NCF (=NF$_{\mbox{ub}}$) &  &  & 2.380 (0.179) &  \\
& NCF (=NF$_{\mbox{tb}}$) &  &  &  & 2.336 (0.186) \\
\hline\hline
JIMA & NF$^{\mbox{x}}_{\mbox{utb}\oplus\mbox{ut}\oplus\mbox{ub}\oplus\mbox{tb}}$ & 0.231 (0.020) & \textbf{0.253 (0.049)} & \textbf{0.248 (0.045)} & \textbf{0.177 (0.040)} \\
\hline\hline
\vspace{1mm}
Ablated JIMA (interactions only) & NF$^{\mbox{x}}_{\mbox{utb}}$ & \textbf{0.195 (0.013)} &  &  &  \\
\vspace{1mm}
& NF$^{\mbox{x}}_{\mbox{ut}}$ &  & 2.353 (0.183) &  &  \\
\vspace{1mm}
& NF$^{\mbox{x}}_{\mbox{ub}}$ &  &  & 2.417 (0.169) &  \\
\vspace{1mm}
& NF$^{\mbox{x}}_{\mbox{tb}}$ &  &  &  & 2.336 (0.186) \\
\hline
Ablated JIMA (joint modeling only) & NF$_{\mbox{utb}\oplus\mbox{ut}}$ & 0.598 (0.743) & 0.536 (0.291) &  &  \\
& NF$_{\mbox{utb}\oplus\mbox{ut}\oplus\mbox{ub}}$ & 0.485 (0.043) & 0.477 (0.038) & 0.513 (0.066) &  \\
& NF$_{\mbox{utb}\oplus\mbox{ut}\oplus\mbox{ub}\oplus\mbox{tb}}$ & 0.497 (0.037) & 0.499 (0.075) & 0.536 (0.077) & 0.476 (0.083) \\
\hline\hline
\end{tabular}
\end{center}
\label{Sim75-100-80}
\end{tiny}
\end{table}


\begin{table}[H]
\begin{tiny}
\caption{Simulated outfit recommendation accuracy comparison between the proposed methods (JIMA) and the competing methods with $N=150$, $T=100$, and $B=80$ using RMSE (with standard deviation in parentheses). The best performance is highlighted in bold.}
\vspace{-4mm}
\begin{center}
\begin{tabular}{cccccc}
\hline
\hline
& & \multicolumn{4}{c}{Data sources for measuring prediction performance} \\
\hline
Model Type & Model Name & utb-tensor & ut-matrix & ub-matrix & tb-matrix \\
\hline
Global Mean Imputation & GMI & 5.513 (0.200) & 2.247 (0.100) & 2.265 (0.097) & 2.246 (0.088) \\
\hline
Tensor/Matrix Factorization & TF (CPD) & 3.067 (0.125) &  &  &  \\
& MF &  & 1.947 (0.094) &  &  \\
& MF &  &  & 2.045 (0.100) &  \\
& MF &  &  &  & 2.121 (0.089) \\
\hline
Deep Learning Benchmarks & NTF (=NF$_{\mbox{utb}}$) & 0.405 (0.040) &  &  &  \\
& NCF (=NF$_{\mbox{ut}}$) &  & 2.108 (0.201) &  &  \\
& NCF (=NF$_{\mbox{ub}}$) &  &  & 2.216 (0.151) &  \\
& NCF (=NF$_{\mbox{tb}}$) &  &  &  & 2.330 (0.141) \\
\hline
\hline
JIMA & NF$^{\mbox{x}}_{\mbox{utb}\oplus\mbox{ut}\oplus\mbox{ub}\oplus\mbox{tb}}$ & 0.194 (0.015) & {\bf 0.212 (0.041)} & {\bf 0.217 (0.050)} & {\bf 0.167 (0.053)} \\
\hline
\hline
\vspace{1mm}
Ablated JIMA (interactions only) & NF$^{\mbox{x}}_{\mbox{utb}}$ & {\bf 0.165 (0.014)} &  &  &  \\
\vspace{1mm}
& NF$^{\mbox{x}}_{\mbox{ut}}$ &  & 2.078 (0.211) &  &  \\
\vspace{1mm}
& NF$^{\mbox{x}}_{\mbox{ub}}$ &  &  & 2.173 (0.199) &  \\
& NF$^{\mbox{x}}_{\mbox{tb}}$ &  &  &  & 2.308 (0.146) \\
\hline
Ablated JIMA (joint modeling only) & NF$_{\mbox{utb}\oplus\mbox{ut}}$ & 0.409 (0.033) & 0.410 (0.043) &  &  \\
& NF$_{\mbox{utb}\oplus\mbox{ut}\oplus\mbox{ub}}$ & 0.413 (0.026) & 0.416 (0.038) & 0.445 (0.056) &  \\
& NF$_{\mbox{utb}\oplus\mbox{ut}\oplus\mbox{ub}\oplus\mbox{tb}}$ & 0.424 (0.040) & 0.450 (0.279) & 0.447 (0.057) & 0.483 (0.084) \\
\hline
\hline
\end{tabular}
\end{center}
\label{Sim150-100-80}
\end{tiny}
\end{table}

\begin{table}[H]
\begin{tiny}
\caption{Simulated outfit recommendation accuracy comparison between the proposed methods (JIMA) and the competing methods with $N=100$, $T=100$, and $B=100$ using MAE (with standard deviation in parentheses). The best performance is highlighted in bold. The computational time (in seconds) is reported as Time.}
\vspace{-4mm}
\begin{center}
\begin{tabular}{ccccccc}
\hline
\hline
& & \multicolumn{4}{c}{Data sources for measuring prediction performance} &\\
\hline
Model Type & Model Name & utb-tensor & ut-matrix & ub-matrix & tb-matrix & Time \\
\hline
Grand Mean Imputation & GMI & 4.208 (0.167) & 1.724 (0.093) & 1.718 (0.076) & 1.702 (0.072) & 0.012 (0.002) \\
\hline
Tensor/Matrix Factorization & TF (CPD) & 2.353 (0.097) &  &  &  & 37.090 (1.939) \\
& MF &  & 1.577 (0.085) &  &  & 0.850 (0.021) \\
& MF &  &  & 1.572 (0.162) &  & 0.846 (0.026) \\
& MF &  &  &  & 1.566 (0.054) & 0.845 (0.021) \\
\hline
Deep Learning Benchmarks & NTF (=NF$_{\mbox{utb}}$) & 0.325 (0.016) &  &  &  & 66.239 (6.174) \\
& NCF (=NF$_{\mbox{ut}}$) &  & 1.731 (0.169) &  &  & 1.266 (0.016) \\
& NCF (=NF$_{\mbox{ub}}$) &  &  & 1.628 (0.023) &  & 1.210 (0.012) \\
& NCF (=NF$_{\mbox{tb}}$) &  &  &  & 1.690 (0.292) & 1.268 (0.041) \\
\hline
\hline
JIMA & NF$^{\mbox{x}}_{\mbox{utb}\oplus\mbox{ut}\oplus\mbox{ub}\oplus\mbox{tb}}$ & 0.153 (0.023) & \textbf{0.164 (0.034)} & \textbf{0.161 (0.030)} & \textbf{0.113 (0.045)} & 219.505 (41.653) \\
\hline
\hline
\vspace{1mm}
Ablated JIMA & NF$^{\mbox{x}}_{\mbox{utb}}$ & \textbf{0.136 (0.007)} &  &  &  & 72.545 (7.613) \\
\vspace{1mm}
(interactions only) & NF$^{\mbox{x}}_{\mbox{ut}}$ &  & 1.659 (0.010) &  &  & 1.274 (0.018) \\
\vspace{1mm}
& NF$^{\mbox{x}}_{\mbox{ub}}$ &  &  & 1.678 (0.186) &  & 1.282 (0.029) \\
\vspace{1mm}
& NF$^{\mbox{x}}_{\mbox{tb}}$ &  &  &  & 1.668 (0.158) & 1.287 (0.003) \\
\hline
Ablated JIMA & NF$_{\mbox{utb}\oplus\mbox{ut}}$ & 0.329 (0.005) & 0.352 (0.015) &  &  & 112.865 (10.174) \\
(joint modeling only) & NF$_{\mbox{utb}\oplus\mbox{ut}\oplus\mbox{ub}}$ & 0.321 (0.025) & 0.348 (0.023) & 0.304 (0.022) &  & 137.717 (39.271) \\
& NF$_{\mbox{utb}\oplus\mbox{ut}\oplus\mbox{ub}\oplus\mbox{tb}}$ & 0.332 (0.012) & 0.351 (0.075) & 0.344 (0.048) & 0.321 (0.046) & 210.602 (35.512) \\
\hline
\hline
\end{tabular}
\end{center}
\label{Sim100-100-100-mae}
\end{tiny}
\end{table}

\begin{table}[H]
\begin{tiny}
\caption{Simulated outfit recommendation accuracy comparison between the proposed methods (JIMA) and the competing methods with $N=50$, $T=25$, and $B=75$ using MAE (with standard deviation in parentheses). The best performance is highlighted in bold. The computational time (in seconds) is reported as Time.}
\vspace{-4mm}
\begin{center}
\begin{tabular}{ccccccc}
\hline
\hline
& & \multicolumn{4}{c}{Data sources for measuring prediction performance} &\\
\hline
Model Type & Model Name & utb-tensor & ut-matrix & ub-matrix & tb-matrix & Time \\
\hline
Global Mean Imputation & GMI & 4.255 (0.259) & 1.738 (0.127) & 1.716 (0.113) & 1.738 (0.126) & 0.002 (0.000) \\
\hline
Tensor/Matrix Factorization & TF (CPD) & 2.452 (0.158) &  &  &  & 10.380 (42.262) \\
& MF &  & 1.717 (0.124) &  &  & 0.500 (0.017) \\
& MF &  &  & 1.683 (0.115) &  & 0.584 (0.018) \\
& MF &  &  &  & 1.714 (0.126) & 0.519 (0.009) \\
\hline
Deep Learning Benchmarks & NTF (=NF$_{\mbox{utb}}$) & 0.679 (0.177) &  &  &  & 11.410 (21.698) \\
& NCF (=NF$_{\mbox{ut}}$) &  & 1.950 (0.126) &  &  & 0.786 (0.028) \\
& NCF (=NF$_{\mbox{ub}}$) &  &  & 1.875 (0.138) &  & 0.913 (0.019) \\
& NCF (=NF$_{\mbox{tb}}$) &  &  &  & 1.984 (0.179) & 0.816 (0.020) \\
\hline
\hline
JIMA & NF$^{\mbox{x}}_{\mbox{utb}\oplus\mbox{ut}\oplus\mbox{ub}\oplus\mbox{tb}}$ & 0.296 (0.045) & \textbf{0.407 (0.106)} & \textbf{0.316 (0.220)} & \textbf{0.333 (0.097)} & 34.941 (85.055) \\
\hline
\hline
\vspace{1mm}
Ablated JIMA & NF$^{\mbox{x}}_{\mbox{utb}}$ & \textbf{0.266 (0.052)} &  &  &  & 8.699 (0.699) \\
\vspace{1mm}
(interactions only) & NF$^{\mbox{x}}_{\mbox{ut}}$ &  & 1.950 (0.139) &  &  & 0.803 (0.018) \\
\vspace{1mm}
& NF$^{\mbox{x}}_{\mbox{ub}}$ &  &  & 1.895 (0.135) &  & 0.930 (0.025) \\
\vspace{1mm}
& NF$^{\mbox{x}}_{\mbox{tb}}$ &  &  &  & 1.985 (0.167) & 0.832 (0.021) \\
\hline
Ablated JIMA & NF$_{\mbox{utb}\oplus\mbox{ut}}$ & 0.671 (0.137) & 0.859 (0.154) &  &  & 12.427 (2.628) \\
(joint modeling only) & NF$_{\mbox{utb}\oplus\mbox{ut}\oplus\mbox{ub}}$ & 0.757 (0.229) & 0.909 (0.181) & 0.637 (0.179) &  & 15.868 (1.573) \\
& NF$_{\mbox{utb}\oplus\mbox{ut}\oplus\mbox{ub}\oplus\mbox{tb}}$ & 0.786 (0.228) & 0.904 (0.140) & 0.668 (0.187) & 0.766 (0.141) & 26.701 (44.807) \\
\hline
\hline
\end{tabular}
\end{center}
\label{Sim50-25-75-mae}
\end{tiny}
\end{table}

\begin{table}[H]
\begin{tiny}
\caption{Simulated outfit recommendation accuracy comparison between the proposed methods (JIMA) and the competing methods with $N=75$, $T=100$, and $B=80$ using MAE (with standard deviation in parentheses). The best performance is highlighted in bold. The computational time used for each method (in seconds) is recorded as Time.}
\vspace{-4mm}
\begin{center}
\begin{tabular}{ccccccc}
\hline\hline
& & \multicolumn{4}{c}{Data sources for measuring prediction performance} &\\
\hline
Model Type & Model Name & utb-tensor & ut-matrix & ub-matrix & tb-matrix & Time \\
\hline
Global Mean Imputation & GMI & 4.201 (0.184) & 1.724 (0.095) & 1.707 (0.085) & 1.703 (0.085) & 0.007 (0.001) \\
\hline
Tensor/Matrix Factorization & TF (CPD) & 2.367 (0.112) &  &  &  & 21.407 (2.302) \\
& MF &  & 1.624 (0.099) &  &  & 0.739 (0.031) \\
& MF &  &  & 1.652 (0.090) &  & 0.677 (0.021) \\
& MF &  &  &  & 1.603 (0.094) & 0.678 (0.024) \\
\hline
Deep Learning Benchmarks & NTF (=NF$_{\mbox{utb}}$) & 0.362 (0.024) &  &  &  & 90.473 (228.322) \\
& NCF (=NF$_{\mbox{ut}}$) &  & 1.776 (0.132) &  &  & 1.133 (0.029) \\
& NCF (=NF$_{\mbox{ub}}$) &  &  & 1.793 (0.135) &  & 1.032 (0.030) \\
& NCF (=NF$_{\mbox{tb}}$) &  &  &  & 1.745 (0.133) & 1.072 (0.042) \\
\hline\hline
\vspace{1mm}
JIMA & NF$^{\mbox{x}}_{\mbox{utb}\oplus\mbox{ut}\oplus\mbox{ub}\oplus\mbox{tb}}$ & 0.178 (0.015) & \textbf{0.187 (0.034)} & \textbf{0.189 (0.031)} & \textbf{0.128 (0.028)} & 122.762 (18.073) \\
\hline\hline
\vspace{1mm}
Ablated JIMA  & NF$^{\mbox{x}}_{\mbox{utb}}$ & \textbf{0.151 (0.009)} &  &  &  & 116.823 (458.527) \\
\vspace{1mm}
(interactions only)& NF$^{\mbox{x}}_{\mbox{ut}}$ &  & 1.755 (0.140) &  &  & 1.137 (0.037) \\
\vspace{1mm}
& NF$^{\mbox{x}}_{\mbox{ub}}$ &  &  & 1.819 (0.125) &  & 1.061 (0.028) \\
\vspace{1mm}
& NF$^{\mbox{x}}_{\mbox{tb}}$ &  &  &  & 1.743 (0.135) & 1.096 (0.040) \\
\hline
Ablated JIMA & NF$^{\mbox{x}}_{\mbox{utb}\oplus\mbox{ut}}$ & 0.461 (0.569) & 0.403 (0.222) &  &  & 62.423 (5.110) \\
(joint modeling only) & NF$^{\mbox{x}}_{\mbox{utb}\oplus\mbox{ut}\oplus\mbox{ub}}$ & 0.374 (0.033) & 0.358 (0.028) & 0.385 (0.047) &  & 90.235 (38.902) \\
& NF$^{\mbox{x}}_{\mbox{utb}\oplus\mbox{ut}\oplus\mbox{ub}\oplus\mbox{tb}}$ & 0.382 (0.030) & 0.375 (0.055) & 0.401 (0.059) & 0.353 (0.061) & 203.092 (516.102) \\
\hline\hline
\end{tabular}
\end{center}
\label{Sim75-100-80-mae}
\end{tiny}
\end{table}

\begin{table}[H]
\begin{tiny}
\caption{Simulated outfit recommendation accuracy comparison between the proposed methods (JIMA) and the competing methods with $N=150$, $T=100$, and $B=80$ using MAE (with standard deviation in parentheses). The best performance is highlighted in bold. The computational time used for each method (in seconds) is recorded as Time.}
\vspace{-4mm}
\begin{center}
\begin{tabular}{ccccccc}
\hline\hline
& & \multicolumn{4}{c}{Data sources for measuring prediction performance} &\\
\hline
Model Type & Model Name & utb-tensor & ut-matrix & ub-matrix & tb-matrix & Time \\
\hline
Global Mean Imputation & GMI & 4.220 (0.156) & 1.708 (0.076) & 1.722 (0.078) & 1.711 (0.069) & 0.046 (0.011) \\
\hline
Tensor/Matrix Factorization & TF (CPD) & 2.306 (0.099) &  &  &  & 86.013 (5.271) \\
& MF &  & 1.444 (0.070) &  &  & 2.144 (0.086) \\
& MF &  &  & 1.526 (0.081) &  & 1.772 (0.060) \\
& MF &  &  &  & 1.599 (0.071) & 1.476 (0.053) \\
\hline
Deep Learning Benchmarks & NTF (=NF$_{\mbox{utb}}$) & 0.312 (0.030) &  &  &  & 197.324 (42.489) \\
& NCF (=NF$_{\mbox{ut}}$) &  & 1.561 (0.155) &  &  & 3.092 (0.111) \\
& NCF (=NF$_{\mbox{ub}}$) &  &  & 1.648 (0.120) &  & 2.710 (0.071) \\
& NCF (=NF$_{\mbox{tb}}$) &  &  &  & 1.742 (0.104) & 2.347 (0.066) \\
\hline\hline
\vspace{1mm}
JIMA & NF$^{\mbox{x}}_{\mbox{utb}\oplus\mbox{ut}\oplus\mbox{ub}\oplus\mbox{tb}}$ & 0.150 (0.011) & \textbf{0.162 (0.029)} & \textbf{0.165 (0.033)} & \textbf{0.121 (0.039)} & 491.183 (64.743) \\
\hline\hline
\vspace{1mm}
Ablated JIMA & NF$^{\mbox{x}}_{\mbox{utb}}$ & \textbf{0.130 (0.010)} &  &  &  & 203.869 (38.665) \\
\vspace{1mm}
(interactions only) & NF$^{\mbox{x}}_{\mbox{ut}}$ &  & 1.534 (0.172) &  &  & 3.134 (0.111) \\
\vspace{1mm}
& NF$^{\mbox{x}}_{\mbox{ub}}$ &  &  & 1.605 (0.152) &  & 2.737 (0.082) \\
\vspace{1mm}
& NF$^{\mbox{x}}_{\mbox{tb}}$ &  &  &  & 1.720 (0.111) & 2.377 (0.072) \\
\hline
Ablated JIMA & NF$_{\mbox{utb}\oplus\mbox{ut}}$ & 0.317 (0.026) & 0.310 (0.034) &  &  & 302.680 (45.548) \\
(joint modeling only) & NF$_{\mbox{utb}\oplus\mbox{ut}\oplus\mbox{ub}}$ & 0.319 (0.020) & 0.312 (0.027) & 0.334 (0.045) &  & 395.969 (58.421) \\
& NF$_{\mbox{utb}\oplus\mbox{ut}\oplus\mbox{ub}\oplus\mbox{tb}}$ & 0.327 (0.031) & 0.338 (0.210) & 0.335 (0.043) & 0.356 (0.060) & 473.904 (64.775) \\
\hline\hline
\end{tabular}
\end{center}
\label{Sim150-100-80-mae}
\end{tiny}
\end{table}

\clearpage

\subsection*{Additional Results and Evaluation of Simulation Study 2}

Next, we provide additional simulation results for Section \ref{sec:sim4d}. The results are measured in root mean square errors (RMSE) in Tables \ref{Sim4d100-20-15-10}, \ref{Sim4d100-10-15-20}, \ref{Sim4d100-15-20-15}, and \ref{Sim4d150-20-20-20}, corresponding to the shapes $(100, 20, 15, 10)$, $(100, 10, 15, 20)$, $(100, 15, 20, 15)$, and $(150,20, 20, 20)$, respectively.
It can be seen that the findings are consistent with the results in the main text. For each method, we also provide all simulation results on mean absolute errors (MAE) and the computational time used. The results are provided in Tables  \ref{Sim4d150-20-30-20-mae}, \ref{Sim4d100-20-15-10-mae}, \ref{Sim4d100-10-15-20-mae}, \ref{Sim4d100-15-20-15-mae}, and \ref{Sim4d150-20-20-20-mae}. It can be seen that the findings are consistent with the RMSE results.

\begin{table}[H]
\begin{tiny}
\caption{Simulated outfit recommendation accuracy comparison between the proposed methods (JIMA) and the competing methods with shape $(100, 20, 15, 10)$ using RMSE (with standard deviation in parentheses). The best performance is highlighted in bold.}
\vspace{-4mm}
\begin{center}
\begin{tabular}{ccccccc}
\hline
\hline
& & \multicolumn{5}{c}{Data sources for measuring prediction performance} \\
\hline
Type & Model Name & utbh-tensor & utb-tensor & ut-matrix & ub-matrix & uh-matrix \\
\hline
Global Mean Imputation & GMI
& 8.583 (0.839) & 5.431 (0.436) & 2.267 (0.164) & 2.223 (0.215) & 2.207 (0.260) \\
\hline

Tensor/Matrix & TF (CPD)
& 4.438 (0.518) &  &  &  &  \\
Factorization & TF (CPD)
&  & 3.718 (0.250) &  &  &  \\
& MF
&  &  & 2.262 (0.172) &  &  \\
& MF
&  &  &  & 2.212 (0.218) &  \\
& MF
&  &  &  &  & 2.234 (0.258) \\
\hline

\vspace{1mm}
Deep Learning & NTF (=NF$_{\mbox{utbh}}$)
& 0.718 (0.146) &  &  &  &  \\
\vspace{1mm}
Benchmarks & NTF (=NF$_{\mbox{utb}}$)
&  & 1.684 (0.316) &  &  &  \\
\vspace{1mm}
& NCF (=NF$_{\mbox{ut}}$)
&  &  & 2.646 (0.199) &  &  \\
\vspace{1mm}
& NCF (=NF$_{\mbox{ub}}$)
&  &  &  & 2.622 (0.304) &  \\
\vspace{1mm}
& NCF (=NF$_{\mbox{uh}}$)
&  &  &  &  & 2.610 (0.308) \\
\hline
\hline
\vspace{1mm}
JIMA & NF$^{\mbox{x}}_{\mbox{utbh}\oplus\mbox{utb}\oplus\mbox{ut}\oplus\mbox{ub}\oplus\mbox{uh}}$
& 0.315 (0.036) & \textbf{0.358 (0.068)} & \textbf{0.389 (0.088)} & \textbf{0.393 (0.081)} & \textbf{0.467 (0.102)} \\
\hline
\hline

\vspace{1mm}
Ablated JIMA & NF$^{\mbox{x}}_{\mbox{utbh}}$
& \textbf{0.271 (0.062)} &  &  &  &  \\
\vspace{1mm}
(interactions only) & NF$^{\mbox{x}}_{\mbox{utb}}$
&  & 0.857 (0.506) &  &  &  \\
\vspace{1mm}
& NF$^{\mbox{x}}_{\mbox{ut}}$
&  &  & 2.643 (0.207) &  &  \\
\vspace{1mm}
& NF$^{\mbox{x}}_{\mbox{ub}}$
&  &  &  & 2.627 (0.278) &  \\
\vspace{1mm}
& NF$^{\mbox{x}}_{\mbox{uh}}$
&  &  &  &  & 2.580 (0.289) \\
\hline

\vspace{1mm}
Ablated JIMA & NF$_{\mbox{utbh}\oplus\mbox{utb}}$
& 0.733 (0.115) & 0.799 (0.092) &  &  &  \\
\vspace{1mm}
(joint modeling only) & NF$_{\mbox{utbh}\oplus\mbox{utb}\oplus\mbox{ut}}$
& 0.702 (0.081) & 0.792 (0.103) & 0.722 (0.099) &  &  \\
\vspace{1mm}
& NF$_{\mbox{utbh}\oplus\mbox{utb}\oplus\mbox{ut}\oplus\mbox{ub}}$
& 0.734 (0.092) & 0.822 (0.104) & 0.707 (0.087) & 0.771 (0.149) &  \\
\vspace{1mm}
& NF$_{\mbox{utbh}\oplus\mbox{utb}\oplus\mbox{ut}\oplus\mbox{ub}\oplus\mbox{uh}}$
& 0.724 (0.085) & 0.827 (0.138) & 0.709 (0.092) & 0.774 (0.105) & 0.907 (0.172) \\
\hline
\hline
\end{tabular}
\end{center}
\label{Sim4d100-20-15-10}
\end{tiny}
\end{table}

\newpage
\begin{table}[H]
\begin{tiny}
\caption{Simulated outfit recommendation accuracy comparison between the proposed methods (JIMA) and the competing methods with shape $(100, 10, 15, 20)$ using RMSE (with standard deviation in parentheses). The best performance is highlighted in bold.}
\vspace{-4mm}
\begin{center}
\begin{tabular}{ccccccc}
\hline
\hline
& & \multicolumn{5}{c}{Data sources for measuring prediction performance} \\
\hline
Type & Model Name & utbh-tensor & utb-tensor & ut-matrix & ub-matrix & uh-matrix \\
\hline
Global Mean Imputation & GMI
& 8.629 (0.731) & 5.489 (0.538) & 2.258 (0.248) & 2.268 (0.178) & 2.218 (0.185) \\
\hline

Tensor/Matrix & TF (CPD)
& 4.401 (0.557) &  &  &  &  \\
Factorization & TF (CPD)
&  & 4.138 (0.438) &  &  &  \\
& MF
&  &  & 2.254 (0.235) &  &  \\
& MF
&  &  &  & 2.258 (0.188) &  \\
& MF
&  &  &  &  & 2.198 (0.189) \\
\hline

\vspace{1mm}
Deep Learning & NTF (=NF$_{\mbox{utbh}}$)
& 0.687 (0.060) &  &  &  &  \\
\vspace{1mm}
Benchmarks & NTF (=NF$_{\mbox{utb}}$)
&  & 2.297 (0.592) &  &  &  \\
\vspace{1mm}
& NCF (=NF$_{\mbox{ut}}$)
&  &  & 2.631 (0.302) &  &  \\
\vspace{1mm}
& NCF (=NF$_{\mbox{ub}}$)
&  &  &  & 2.646 (0.225) &  \\
\vspace{1mm}
& NCF (=NF$_{\mbox{uh}}$)
&  &  &  &  & 2.580 (0.234) \\
\hline
\hline

\vspace{1mm}
JIMA & NF$^{\mbox{x}}_{\mbox{utbh}\oplus\mbox{utb}\oplus\mbox{ut}\oplus\mbox{ub}\oplus\mbox{uh}}$
& 0.346 (0.057) & \textbf{0.463 (0.136)} & \textbf{0.473 (0.078)} & \textbf{0.435 (0.104)} & \textbf{0.377 (0.066)} \\
\hline
\hline

\vspace{1mm}
Ablated JIMA & NF$^{\mbox{x}}_{\mbox{utbh}}$
& \textbf{0.278 (0.056)} &  &  &  &  \\
\vspace{1mm}
(interactions only) & NF$^{\mbox{x}}_{\mbox{utb}}$
&  & 1.795 (0.580) &  &  &  \\
\vspace{1mm}
& NF$^{\mbox{x}}_{\mbox{ut}}$
&  &  & 2.639 (0.297) &  &  \\
\vspace{1mm}
& NF$^{\mbox{x}}_{\mbox{ub}}$
&  &  &  & 2.649 (0.221) &  \\
\vspace{1mm}
& NF$^{\mbox{x}}_{\mbox{uh}}$
&  &  &  &  & 2.599 (0.249) \\
\hline

\vspace{1mm}
Ablated JIMA & NF$_{\mbox{utbh}\oplus\mbox{utb}}$
& 0.725 (0.089) & 0.886 (0.119) &  &  &  \\
\vspace{1mm}
(joint modeling only) & NF$_{\mbox{utbh}\oplus\mbox{utb}\oplus\mbox{ut}}$
& 0.720 (0.090) & 0.885 (0.102) & 0.880 (0.126) &  &  \\
\vspace{1mm}
& NF$_{\mbox{utbh}\oplus\mbox{utb}\oplus\mbox{ut}\oplus\mbox{ub}}$
& 0.730 (0.079) & 0.914 (0.146) & 0.897 (0.185) & 0.845 (0.286) &  \\
\vspace{1mm}
& NF$_{\mbox{utbh}\oplus\mbox{utb}\oplus\mbox{ut}\oplus\mbox{ub}\oplus\mbox{uh}}$
& 0.739 (0.096) & 0.923 (0.131) & 0.892 (0.156) & 0.803 (0.131) & 0.686 (0.098) \\
\hline
\hline
\end{tabular}
\end{center}
\label{Sim4d100-10-15-20}
\end{tiny}
\end{table}

\begin{table}[H]
\begin{tiny}
\caption{Simulated outfit recommendation accuracy comparison between the proposed methods (JIMA) and the competing methods with shape $(100, 15, 20, 15)$ using RMSE (with standard deviation in parentheses). The best performance is highlighted in bold.}
\vspace{-4mm}
\begin{center}
\begin{tabular}{ccccccc}
\hline
\hline
& & \multicolumn{5}{c}{Data sources for measuring prediction performance} \\
\hline
Type & Model Name & utbh-tensor & utb-tensor & ut-matrix & ub-matrix & uh-matrix \\
\hline
Global Mean Imputation & GMI
& 8.633 (0.702) & 5.464 (0.436) & 2.266 (0.193) & 2.241 (0.189) & 2.234 (0.198) \\
\hline

Tensor/Matrix & TF (CPD)
& 4.284 (0.586) &  &  &  &  \\
Factorization & TF (CPD)
&  & 3.765 (0.318) &  &  &  \\
& MF
&  &  & 2.292 (0.194) &  &  \\
& MF
&  &  &  & 2.242 (0.201) &  \\
& MF
&  &  &  &  & 2.258 (0.188) \\
\hline

\vspace{1mm}
Deep Learning & NTF (=NF$_{\mbox{utbh}}$)
& 0.642 (0.066) &  &  &  &  \\
\vspace{1mm}
Benchmarks & NTF (=NF$_{\mbox{utb}}$)
&  & 1.763 (0.521) &  &  &  \\
\vspace{1mm}
& NCF (=NF$_{\mbox{ut}}$)
&  &  & 2.688 (0.208) &  &  \\
\vspace{1mm}
& NCF (=NF$_{\mbox{ub}}$)
&  &  &  & 2.637 (0.228) &  \\
\vspace{1mm}
& NCF (=NF$_{\mbox{uh}}$)
&  &  &  &  & 2.657 (0.255) \\
\hline
\hline

\vspace{1mm}
JIMA & NF$^{\mbox{x}}_{\mbox{utbh}\oplus\mbox{utb}\oplus\mbox{ut}\oplus\mbox{ub}\oplus\mbox{uh}}$
& 0.308 (0.039) & \textbf{0.362 (0.070)} & \textbf{0.418 (0.105)} & \textbf{0.374 (0.081)} & \textbf{0.407 (0.081)} \\
\hline
\hline

\vspace{1mm}
Ablated JIMA & NF$^{\mbox{x}}_{\mbox{utbh}}$
& \textbf{0.251 (0.036)} &  &  &  &  \\
\vspace{1mm}
(interactions only) & NF$^{\mbox{x}}_{\mbox{utb}}$
&  & 0.810 (0.350) &  &  &  \\
\vspace{1mm}
& NF$^{\mbox{x}}_{\mbox{ut}}$
&  &  & 2.664 (0.208) &  &  \\
\vspace{1mm}
& NF$^{\mbox{x}}_{\mbox{ub}}$
&  &  &  & 2.642 (0.239) &  \\
\vspace{1mm}
& NF$^{\mbox{x}}_{\mbox{uh}}$
&  &  &  &  & 2.601 (0.266) \\
\hline

\vspace{1mm}
Ablated JIMA & NF$_{\mbox{utbh}\oplus\mbox{utb}}$
& 0.683 (0.076) & 0.811 (0.072) &  &  &  \\
\vspace{1mm}
(joint modeling only) & NF$_{\mbox{utbh}\oplus\mbox{utb}\oplus\mbox{ut}}$
& 0.668 (0.075) & 0.796 (0.122) & 0.803 (0.164) &  &  \\
\vspace{1mm}
& NF$_{\mbox{utbh}\oplus\mbox{utb}\oplus\mbox{ut}\oplus\mbox{ub}}$
& 0.686 (0.082) & 0.836 (0.125) & 0.803 (0.132) & 0.708 (0.113) &  \\
\vspace{1mm}
& NF$_{\mbox{utbh}\oplus\mbox{utb}\oplus\mbox{ut}\oplus\mbox{ub}\oplus\mbox{uh}}$
& 0.671 (0.071) & 0.793 (0.095) & 0.797 (0.100) & 0.717 (0.104) & 0.799 (0.117) \\
\hline
\hline
\end{tabular}
\end{center}
\label{Sim4d100-15-20-15}
\end{tiny}
\end{table}

\begin{table}[H]
\begin{tiny}
\caption{Simulated outfit recommendation accuracy comparison between the proposed methods (JIMA) and the competing methods with shape $(150, 20, 20, 20)$ using RMSE (with standard deviation in parentheses). The best performance is highlighted in bold.}
\vspace{-4mm}
\begin{center}
\begin{tabular}{ccccccc}
\hline
\hline
& & \multicolumn{5}{c}{Data sources for measuring prediction performance} \\
\hline
Type & Model Name & utbh-tensor & utb-tensor & ut-matrix & ub-matrix & uh-matrix \\
\hline
Global Mean Imputation & GMI
& 8.726 (0.713) & 5.593 (0.471) & 2.277 (0.180) & 2.275 (0.166) & 2.213 (0.207) \\
\hline
Tensor/Matrix & TF (CPD)
& 4.077 (0.462) & & & & \\
Factorization & TF (CPD)
& & 3.379 (0.278) & & & \\
& MF
& & & 2.259 (0.193) & & \\
& MF
& & & & 2.228 (0.159) & \\
& MF
& & & & & 2.241 (0.210) \\
\hline
\vspace{1mm}
Deep Learning & NTF (=NF$_{\mbox{utbh}}$)
& 0.497 (0.062) & & & & \\
\vspace{1mm}
Benchmarks & NTF (=NF$_{\mbox{utb}}$)
& & 1.090 (0.339) & & & \\
\vspace{1mm}
& NCF (=NF$_{\mbox{ut}}$)
& & & 2.694 (0.236) & & \\
\vspace{1mm}
& NCF (=NF$_{\mbox{ub}}$)
& & & & 2.636 (0.268) & \\
\vspace{1mm}
& NCF (=NF$_{\mbox{uh}}$)
& & & & & 2.654 (0.321) \\
\hline
\hline
\vspace{1mm}
JIMA & NF$^{\mbox{x}}_{\mbox{utbh}\oplus\mbox{utb}\oplus\mbox{ut}\oplus\mbox{ub}\oplus\mbox{uh}}$
& 0.244 (0.026) & \textbf{0.276 (0.051)} & \textbf{0.338 (0.304)} & \textbf{0.300 (0.069)} & \textbf{0.306 (0.082)} \\
\hline
\hline
\vspace{1mm}
Ablated JIMA & NF$^{\mbox{x}}_{\mbox{utbh}}$
& \textbf{0.194 (0.017)} & & & & \\
\vspace{1mm}
(interactions only) & NF$^{\mbox{x}}_{\mbox{utb}}$
& & 0.564 (0.433) & & & \\
\vspace{1mm}
& NF$^{\mbox{x}}_{\mbox{ut}}$
& & & 2.682 (0.247) & & \\
\vspace{1mm}
& NF$^{\mbox{x}}_{\mbox{ub}}$
& & & & 2.634 (0.270) & \\
\vspace{1mm}
& NF$^{\mbox{x}}_{\mbox{uh}}$
& & & & & 2.666 (0.256) \\
\hline
\vspace{1mm}
Ablated JIMA & NF$_{\mbox{utbh}\oplus\mbox{utb}}$
& 0.535 (0.061) & 0.688 (0.081) & & & \\
\vspace{1mm}
(joint modeling only) & NF$_{\mbox{utbh}\oplus\mbox{utb}\oplus\mbox{ut}}$
& 0.522 (0.049) & 0.666 (0.070) & 0.614 (0.072) & & \\
\vspace{1mm}
& NF$_{\mbox{utbh}\oplus\mbox{utb}\oplus\mbox{ut}\oplus\mbox{ub}}$
& 0.516 (0.050) & 0.685 (0.098) & 0.628 (0.076) & 0.623 (0.100) & \\
\vspace{1mm}
& NF$_{\mbox{utbh}\oplus\mbox{utb}\oplus\mbox{ut}\oplus\mbox{ub}\oplus\mbox{uh}}$
& 0.517 (0.054) & 0.650 (0.086) & 0.603 (0.083) & 0.599 (0.068) & 0.615 (0.088) \\
\hline
\hline
\end{tabular}
\end{center}
\label{Sim4d150-20-20-20}
\end{tiny}
\end{table}

\begin{table}[H]
\begin{tiny}
\caption{Simulated outfit recommendation accuracy comparison between the proposed methods (JIMA) and the competing methods with shape $(150,20,30,20)$ using MAE (with standard deviation in parentheses). The best performance is highlighted in bold.}
\vspace{-4mm}
\begin{center}
\begin{tabular}{ccccccc}
\hline
\hline
 & \multicolumn{5}{c}{Data sources for measuring prediction performance} \\
\hline
 Model Name & utbh-tensor & utb-tensor & ut-matrix & ub-matrix & uh-matrix & Time \\
\hline
 GMI 
& 6.718 (0.527) & 4.290 (0.361) & 1.737 (0.143) & 1.732 (0.113) & 1.694 (0.141) & 0.025 (0.002) \\
\hline
 TF (CPD) 
& 3.080 (0.302) &  &  &  &  & 69.700 (4.814) \\
 TF (CPD) 
&  & 2.490 (0.192) &  &  &  & 4.075 (0.224) \\
 MF 
&  &  & 1.727 (0.150) &  &  & 0.655 (0.026) \\
 MF 
&  &  &  & 1.668 (0.108) &  & 0.648 (0.021) \\
 MF 
&  &  &  &  & 1.628 (0.105) & 0.574 (0.022) \\
\hline
 NTF (=NF$_{\mbox{utbh}}$) 
& 0.333 (0.033) &  &  &  &  & 129.303 (8.268) \\
 NTF (=NF$_{\mbox{utb}}$)
&  & 0.670 (0.128) &  &  &  & 7.721 (0.562) \\
 NCF (=NF$_{\mbox{ut}}$) 
&  &  & 2.038 (0.211) &  &  & 1.009 (0.042) \\
 NCF (=NF$_{\mbox{ub}}$) 
&  &  &  & 1.952 (0.151) &  & 1.148 (0.054) \\
 NCF (=NF$_{\mbox{uh}}$) 
&  &  &  &  & 1.997 (0.199) & 1.012 (0.042) \\
\hline
\hline
\vspace{1mm}
NF$^{\mbox{x}}_{\mbox{utbh}\oplus\mbox{utb}\oplus\mbox{ut}\oplus\mbox{ub}\oplus\mbox{uh}}$ 
& 0.163 (0.013) & \textbf{0.185 (0.029)} & \textbf{0.232 (0.065)} & \textbf{0.202 (0.037)} & \textbf{0.224 (0.062)} & 607.648 (62.991) \\
\hline
\hline
\vspace{1mm}
 NF$^{\mbox{x}}_{\mbox{utbh}}$ 
& \textbf{0.133 (0.008)} &  &  &  &  & 161.065 (13.663) \\
\vspace{1mm}
 NF$^{\mbox{x}}_{\mbox{utb}}$ 
&  & 0.309 (0.085) &  &  &  & 8.365 (0.478) \\
\vspace{1mm}
 NF$^{\mbox{x}}_{\mbox{ut}}$ 
&  &  & 2.046 (0.212) &  &  & 1.016 (0.036) \\
\vspace{1mm}
 NF$^{\mbox{x}}_{\mbox{ub}}$ 
&  &  &  & 1.932 (0.186) &  & 1.132 (0.042) \\
\vspace{1mm}
 NF$^{\mbox{x}}_{\mbox{uh}}$ 
&  &  &  &  & 2.002 (0.203) & 1.019 (0.032) \\
\hline
 NF$_{\mbox{utbh}\oplus \mbox{utb}}$ 
& 0.347 (0.036) & 0.470 (0.064) &  &  &  & 237.510 (32.085) \\
 NF$_{\mbox{utbh}\oplus \mbox{utb} \oplus \mbox{ut}}$ 
& 0.368 (0.033) & 0.477 (0.052) & 0.449 (0.062) &  &  & 383.630 (382.826) \\
 NF$_{\mbox{utbh}\oplus \mbox{utb} \oplus \mbox{ut} \oplus \mbox{ub}}$ 
& 0.360 (0.034) & 0.481 (0.057) & 0.461 (0.057) & 0.409 (0.055) &  & 382.184 (69.673) \\
 NF$_{\mbox{utbh}\oplus \mbox{utb} \oplus \mbox{ut} \oplus \mbox{ub} \oplus \mbox{uh}}$ 
& 0.364 (0.037) & 0.472 (0.049) & 0.465 (0.068) & 0.467 (0.282) & 0.453 (0.088) & 473.948 (52.790) \\
\hline
\hline
\end{tabular}
\end{center}
\label{Sim4d150-20-30-20-mae}
\end{tiny}
\end{table}

\begin{table}[H]
\begin{tiny}
\caption{Simulated outfit recommendation accuracy comparison between the proposed methods (JIMA) and the competing methods with shape $(100,20,15,10)$ using MAE (with standard deviation in parentheses). The best performance is highlighted in bold.}
\vspace{-4mm}
\begin{center}
\begin{tabular}{ccccccc}
\hline
\hline
 & \multicolumn{5}{c}{Data sources for measuring prediction performance} \\
\hline
 Model Name & utbh-tensor & utb-tensor & ut-matrix & ub-matrix & uh-matrix & Time \\
\hline
 GMI 
& 6.644 (0.641) & 4.183 (0.344) & 1.731 (0.133) & 1.696 (0.169) & 1.691 (0.205) & 0.006 (0.001) \\
\hline
 TF (CPD) 
& 3.366 (0.418) &  &  &  &  & 12.184 (0.670) \\
 TF (CPD) 
&  & 2.834 (0.195) &  &  &  & 1.699 (0.069) \\
 MF 
&  &  & 1.727 (0.139) &  &  & 0.529 (0.013) \\
 MF 
&  &  &  & 1.687 (0.173) &  & 0.505 (0.013) \\
 MF 
&  &  &  &  & 1.715 (0.199) & 0.482 (0.010) \\
\hline
 NTF (=NF$_{\mbox{utbh}}$) 
& 0.548 (0.066) &  &  &  &  & 22.467 (1.494) \\
 NTF (=NF$_{\mbox{utb}}$)
&  & 1.269 (0.237) &  &  &  & 3.183 (0.203) \\
 NCF (=NF$_{\mbox{ut}}$) 
&  &  & 2.031 (0.153) &  &  & 0.911 (0.028) \\
 NCF (=NF$_{\mbox{ub}}$) 
&  &  &  & 2.018 (0.240) &  & 0.881 (0.032) \\
 NCF (=NF$_{\mbox{uh}}$) 
&  &  &  &  & 2.021 (0.239) & 0.840 (0.026) \\
\hline
\hline
\vspace{1mm}
NF$^{\mbox{x}}_{\mbox{utbh}\oplus\mbox{utb}\oplus\mbox{ut}\oplus\mbox{ub}\oplus\mbox{uh}}$ 
& 0.244 (0.029) & \textbf{0.275 (0.053)} & \textbf{0.285 (0.065)} & \textbf{0.289 (0.060)} & \textbf{0.351 (0.072)} & 108.564 (10.985) \\
\hline
\hline
\vspace{1mm}
 NF$^{\mbox{x}}_{\mbox{utbh}}$ 
& \textbf{0.208 (0.044)} &  &  &  &  & 27.751 (1.858) \\
\vspace{1mm}
 NF$^{\mbox{x}}_{\mbox{utb}}$ 
&  & 0.636 (0.346) &  &  &  & 3.451 (0.247) \\
\vspace{1mm}
 NF$^{\mbox{x}}_{\mbox{ut}}$ 
&  &  & 2.025 (0.163) &  &  & 0.924 (0.024) \\
\vspace{1mm}
 NF$^{\mbox{x}}_{\mbox{ub}}$ 
&  &  &  & 2.020 (0.228) &  & 0.891 (0.023) \\
\vspace{1mm}
 NF$^{\mbox{x}}_{\mbox{uh}}$ 
&  &  &  &  & 1.990 (0.229) & 0.850 (0.020) \\
\hline
 NF$_{\mbox{utbh}\oplus \mbox{utb}}$ 
& 0.554 (0.061) & 0.617 (0.072) &  &  &  & 40.577 (4.016) \\
 NF$_{\mbox{utbh}\oplus \mbox{utb} \oplus \mbox{ut}}$ 
& 0.540 (0.056) & 0.607 (0.070) & 0.538 (0.073) &  &  & 52.685 (4.644) \\
 NF$_{\mbox{utbh}\oplus \mbox{utb} \oplus \mbox{ut} \oplus \mbox{ub}}$ 
& 0.571 (0.077) & 0.627 (0.071) & 0.529 (0.064) & 0.575 (0.114) &  & 68.413 (8.659) \\
 NF$_{\mbox{utbh}\oplus \mbox{utb} \oplus \mbox{ut} \oplus \mbox{ub} \oplus \mbox{uh}}$ 
& 0.560 (0.062) & 0.624 (0.074) & 0.529 (0.070) & 0.574 (0.077) & 0.665 (0.131) & 85.749 (8.337) \\
\hline
\hline
\end{tabular}
\end{center}
\label{Sim4d100-20-15-10-mae}
\end{tiny}
\end{table}

\begin{table}[H]
\begin{tiny}
\caption{Simulated outfit recommendation accuracy comparison between the proposed methods (JIMA) and the competing methods with shape $(100,10,15,20)$ using MAE (with standard deviation in parentheses). The best performance is highlighted in bold.}
\vspace{-4mm}
\begin{center}
\begin{tabular}{ccccccc}
\hline
\hline
 & \multicolumn{5}{c}{Data sources for measuring prediction performance} \\
\hline
 Model Name & utbh-tensor & utb-tensor & ut-matrix & ub-matrix & uh-matrix & Time \\
\hline
 GMI 
& 6.664 (0.579) & 4.235 (0.429) & 1.734 (0.191) & 1.735 (0.144) & 1.699 (0.144) & 0.006 (0.001) \\
\hline
 TF (CPD) 
& 3.348 (0.445) &  &  &  &  & 11.985 (0.965) \\
 TF (CPD) 
&  & 3.148 (0.343) &  &  &  & 1.109 (0.056) \\
 MF 
&  &  & 1.732 (0.184) &  &  & 0.493 (0.016) \\
 MF 
&  &  &  & 1.726 (0.153) &  & 0.508 (0.019) \\
 MF 
&  &  &  &  & 1.681 (0.144) & 0.523 (0.024) \\
\hline
 NTF (=NF$_{\mbox{utbh}}$) 
& 0.534 (0.048) &  &  &  &  & 23.085 (1.740) \\
 NTF (=NF$_{\mbox{utb}}$)
&  & 1.736 (0.449) &  &  &  & 2.073 (0.141) \\
 NCF (=NF$_{\mbox{ut}}$) 
&  &  & 2.035 (0.234) &  &  & 0.856 (0.033) \\
 NCF (=NF$_{\mbox{ub}}$) 
&  &  &  & 2.028 (0.180) &  & 0.899 (0.035) \\
 NCF (=NF$_{\mbox{uh}}$) 
&  &  &  &  & 1.978 (0.182) & 0.933 (0.037) \\
\hline
\hline
NF$^{\mbox{x}}_{\mbox{utbh}\oplus\mbox{utb}\oplus\mbox{ut}\oplus\mbox{ub}\oplus\mbox{uh}}$ 
& 0.268 (0.041) & \textbf{0.355 (0.102)} & \textbf{0.350 (0.054)} & \textbf{0.321 (0.073)} & \textbf{0.287 (0.055)} & 99.298 (13.088) \\
\hline
\hline
 NF$^{\mbox{x}}_{\mbox{utbh}}$ 
& \textbf{0.213 (0.034)} &  &  &  &  & 28.232 (1.905) \\
 NF$^{\mbox{x}}_{\mbox{utb}}$ 
&  & 1.334 (0.429) &  &  &  & 2.212 (0.152) \\
 NF$^{\mbox{x}}_{\mbox{ut}}$ 
&  &  & 2.041 (0.240) &  &  & 0.862 (0.034) \\
 NF$^{\mbox{x}}_{\mbox{ub}}$ 
&  &  &  & 2.028 (0.176) &  & 0.909 (0.034) \\
 NF$^{\mbox{x}}_{\mbox{uh}}$ 
&  &  &  &  & 1.995 (0.203) & 0.940 (0.034) \\
\hline
 NF$_{\mbox{utbh}\oplus \mbox{utb}}$ 
& 0.562 (0.065) & 0.677 (0.087) &  &  &  & 40.530 (4.211) \\
 NF$_{\mbox{utbh}\oplus \mbox{utb} \oplus \mbox{ut}}$ 
& 0.554 (0.062) & 0.682 (0.081) & 0.657 (0.085) &  &  & 49.757 (6.363) \\
 NF$_{\mbox{utbh}\oplus \mbox{utb} \oplus \mbox{ut} \oplus \mbox{ub}}$ 
& 0.568 (0.063) & 0.704 (0.113) & 0.660 (0.133) & 0.630 (0.222) &  & 65.168 (8.943) \\
 NF$_{\mbox{utbh}\oplus \mbox{utb} \oplus \mbox{ut} \oplus \mbox{ub} \oplus \mbox{uh}}$ 
& 0.572 (0.066) & 0.705 (0.094) & 0.662 (0.118) & 0.601 (0.109) & 0.513 (0.072) & 81.143 (12.668) \\
\hline
\hline
\end{tabular}
\end{center}
\label{Sim4d100-10-15-20-mae}
\end{tiny}
\end{table}

\begin{table}[H]
\begin{tiny}
\caption{Simulated outfit recommendation accuracy comparison between the proposed methods (JIMA) and the competing methods with shape $(100,15,20,15)$ using MAE (with standard deviation in parentheses). The best performance is highlighted in bold.}
\vspace{-4mm}
\begin{center}
\begin{tabular}{ccccccc}
\hline
\hline
 & \multicolumn{5}{c}{Data sources for measuring prediction performance} \\
\hline
 Model Name & utbh-tensor & utb-tensor & ut-matrix & ub-matrix & uh-matrix & Time \\
\hline
 GMI 
& 6.645 (0.560) & 4.197 (0.353) & 1.730 (0.158) & 1.712 (0.150) & 1.700 (0.159) & 0.007 (0.001) \\
\hline
 TF (CPD) 
& 3.241 (0.459) &  &  &  &  & 18.815 (1.143) \\
 TF (CPD) 
&  & 2.855 (0.256) &  &  &  & 1.626 (0.112) \\
 MF 
&  &  & 1.749 (0.161) &  &  & 0.510 (0.013) \\
 MF 
&  &  &  & 1.704 (0.161) &  & 0.514 (0.016) \\
 MF 
&  &  &  &  & 1.719 (0.153) & 0.502 (0.017) \\
\hline
 NTF (=NF$_{\mbox{utbh}}$) 
& 0.496 (0.053) &  &  &  &  & 33.365 (2.120) \\
 NTF (=NF$_{\mbox{utb}}$)
&  & 1.308 (0.326) &  &  &  & 2.969 (0.190) \\
 NCF (=NF$_{\mbox{ut}}$) 
&  &  & 2.066 (0.167) &  &  & 0.869 (0.024) \\
 NCF (=NF$_{\mbox{ub}}$) 
&  &  &  & 2.015 (0.183) &  & 0.900 (0.028) \\
 NCF (=NF$_{\mbox{uh}}$) 
&  &  &  &  & 2.039 (0.204) & 0.873 (0.021) \\
\hline
\hline
\vspace{1mm}
NF$^{\mbox{x}}_{\mbox{utbh}\oplus\mbox{utb}\oplus\mbox{ut}\oplus\mbox{ub}\oplus\mbox{uh}}$ 
& 0.238 (0.030) & \textbf{0.275 (0.051)} & \textbf{0.314 (0.081)} & \textbf{0.281 (0.067)} & \textbf{0.301 (0.063)} & 139.388 (17.235) \\
\hline
\hline
\vspace{1mm}
 NF$^{\mbox{x}}_{\mbox{utbh}}$ 
& \textbf{0.194 (0.030)} &  &  &  &  & 42.386 (1.990) \\
\vspace{1mm}
 NF$^{\mbox{x}}_{\mbox{utb}}$ 
&  & 0.608 (0.258) &  &  &  & 3.229 (0.164) \\
\vspace{1mm}
 NF$^{\mbox{x}}_{\mbox{ut}}$ 
&  &  & 2.043 (0.172) &  &  & 0.885 (0.025) \\
\vspace{1mm}
 NF$^{\mbox{x}}_{\mbox{ub}}$ 
&  &  &  & 2.017 (0.193) &  & 0.914 (0.031) \\
\vspace{1mm}
 NF$^{\mbox{x}}_{\mbox{uh}}$ 
&  &  &  &  & 1.990 (0.212) & 0.881 (0.025) \\
\hline
 NF$_{\mbox{utbh}\oplus \mbox{utb}}$ 
& 0.530 (0.062) & 0.622 (0.058) &  &  &  & 49.006 (5.040) \\
 NF$_{\mbox{utbh}\oplus \mbox{utb} \oplus \mbox{ut}}$ 
& 0.516 (0.058) & 0.608 (0.095) & 0.597 (0.113) &  &  & 68.702 (11.245) \\
 NF$_{\mbox{utbh}\oplus \mbox{utb} \oplus \mbox{ut} \oplus \mbox{ub}}$ 
& 0.532 (0.064) & 0.639 (0.094) & 0.596 (0.094) & 0.532 (0.095) &  & 88.842 (11.364) \\
 NF$_{\mbox{utbh}\oplus \mbox{utb} \oplus \mbox{ut} \oplus \mbox{ub} \oplus \mbox{uh}}$ 
& 0.519 (0.056) & 0.607 (0.074) & 0.593 (0.075) & 0.542 (0.088) & 0.590 (0.086) & 106.175 (14.478) \\
\hline
\hline
\end{tabular}
\end{center}
\label{Sim4d100-15-20-15-mae}
\end{tiny}
\end{table}

\begin{table}[H]
\begin{tiny}
\caption{Simulated outfit recommendation accuracy comparison between the proposed methods (JIMA) and the competing methods with shape $(150,20,20,20)$ using MAE (with standard deviation in parentheses). The best performance is highlighted in bold.}
\vspace{-4mm}
\begin{center}
\begin{tabular}{ccccccc}
\hline
\hline
 & \multicolumn{5}{c}{Data sources for measuring prediction performance} \\
\hline
 Model Name & utbh-tensor & utb-tensor & ut-matrix & ub-matrix & uh-matrix & Time \\
\hline
 GMI 
& 6.734 (0.566) & 4.302 (0.380) & 1.738 (0.143) & 1.738 (0.130) & 1.684 (0.157) & 0.016 (0.003) \\
\hline
 TF (CPD) 
& 3.072 (0.367) &  &  &  &  & 48.167 (2.632) \\
 TF (CPD) 
&  & 2.570 (0.227) &  &  &  & 2.704 (0.132) \\
 MF 
&  &  & 1.719 (0.150) &  &  & 0.582 (0.014) \\
 MF 
&  &  &  & 1.705 (0.129) &  & 0.571 (0.012) \\
 MF 
&  &  &  &  & 1.700 (0.161) & 0.569 (0.013) \\
\hline
 NTF (=NF$_{\mbox{utbh}}$) 
& 0.382 (0.048) &  &  &  &  & 89.896 (7.047) \\
 NTF (=NF$_{\mbox{utb}}$)
&  & 0.832 (0.250) &  &  &  & 5.333 (0.450) \\
 NCF (=NF$_{\mbox{ut}}$) 
&  &  & 2.038 (0.183) &  &  & 1.007 (0.027) \\
 NCF (=NF$_{\mbox{ub}}$) 
&  &  &  & 2.007 (0.201) &  & 1.027 (0.036) \\
 NCF (=NF$_{\mbox{uh}}$) 
&  &  &  &  & 2.010 (0.249) & 1.014 (0.033) \\
\hline
\hline
\vspace{1mm}
NF$^{\mbox{x}}_{\mbox{utbh}\oplus\mbox{utb}\oplus\mbox{ut}\oplus\mbox{ub}\oplus\mbox{uh}}$ 
& 0.189 (0.020) & \textbf{0.211 (0.038)} & \textbf{0.257 (0.229)} & \textbf{0.226 (0.051)} & \textbf{0.229 (0.058)} & 422.357 (200.340) \\
\hline
\hline
\vspace{1mm}
 NF$^{\mbox{x}}_{\mbox{utbh}}$ 
& \textbf{0.151 (0.011)} &  &  &  &  & 107.330 (8.299) \\
\vspace{1mm}
 NF$^{\mbox{x}}_{\mbox{utb}}$ 
&  & 0.396 (0.195) &  &  &  & 5.638 (0.536) \\
\vspace{1mm}
 NF$^{\mbox{x}}_{\mbox{ut}}$ 
&  &  & 2.031 (0.190) &  &  & 1.025 (0.032) \\
\vspace{1mm}
 NF$^{\mbox{x}}_{\mbox{ub}}$ 
&  &  &  & 1.991 (0.200) &  & 1.023 (0.032) \\
\vspace{1mm}
 NF$^{\mbox{x}}_{\mbox{uh}}$ 
&  &  &  &  & 2.014 (0.195) & 1.031 (0.040) \\
\hline
 NF$_{\mbox{utbh}\oplus \mbox{utb}}$ 
& 0.413 (0.047) & 0.527 (0.059) &  &  &  & 192.599 (178.383) \\
 NF$_{\mbox{utbh}\oplus \mbox{utb} \oplus \mbox{ut}}$ 
& 0.402 (0.037) & 0.511 (0.054) & 0.455 (0.055) &  &  & 425.844 (798.320) \\
 NF$_{\mbox{utbh}\oplus \mbox{utb} \oplus \mbox{ut} \oplus \mbox{ub}}$ 
& 0.396 (0.037) & 0.523 (0.070) & 0.468 (0.056) & 0.469 (0.081) &  & 289.042 (167.675) \\
 NF$_{\mbox{utbh}\oplus \mbox{utb} \oplus \mbox{ut} \oplus \mbox{ub} \oplus \mbox{uh}}$ 
& 0.400 (0.042) & 0.501 (0.069) & 0.449 (0.063) & 0.449 (0.058) & 0.459 (0.064) & 335.103 (57.895) \\
\hline
\hline
\end{tabular}
\end{center}
\label{Sim4d150-20-20-20-mae}
\end{tiny}
\end{table}

\newpage
\subsection*{Additional Evaluation of Offline Experiments}

We provide the results of offline experiments in mean absolute errors (MAE) as well as the computational time used for each method, corresponding to Section \ref{sec:offline}. The results are provided in Table \ref{JIM_results-mae}. It can be seen that the findings are consistent with the RMSE results.

\begin{table}[H]
\begin{tiny}
\caption{Outfit recommendation accuracy comparison between the proposed method (JIMA) and the competing methods using real offline fashion data. MAEs are reported (with standard deviations in parentheses). The best performance is highlighted in bold. Computational time (in seconds) is reported as Time.}
\vspace{-4mm}
\begin{center}
\begin{tabular}{ccccccc}
\hline
\hline
& & \multicolumn{4}{c}{Data sources for measuring prediction performance} &\\
\hline
Model Type & Model Name & utb-tensor & ut-matrix & ub-matrix & tb-matrix & Time \\
\hline
Grand Mean Imputation & GMI & 0.938 (0.011) & 0.958 (0.017) & 1.019 (0.021) & 0.810 (0.019) & 0.002 (0.000) \\
\hline
Tensor/Matrix Factorization & TF (CPD) & 2.255 (0.320) &  &  &  & 1.917 (0.027) \\
& MF &  & 0.953 (0.052) &  &  & 0.688 (0.014) \\
& MF &  &  & 1.046 (0.066) &  & 0.701 (0.051) \\
& MF &  &  &  & 0.670 (0.018) & 0.603 (0.014) \\
\hline
Deep Learning Benchmarks & NTF (=NF$_{\mbox{utb}}$) & 0.720 (0.034) &  &  &  & 3.127 (0.076) \\
& NCF (=NF$_{\mbox{ut}}$) &  & 0.798 (0.019) &  &  & 1.112 (0.066) \\
& NCF (=NF$_{\mbox{ub}}$) &  &  & 0.841 (0.023) &  & 1.038 (0.018) \\
& NCF (=NF$_{\mbox{tb}}$) &  &  &  & 0.521 (0.016) & 0.984 (0.053) \\
\hline
\hline
JIMA & NF$^{\mbox{x}}_{\mbox{utb}\oplus\mbox{ut}\oplus\mbox{ub}\oplus\mbox{tb}}$ & {\bf 0.701 (0.015)} & {\bf 0.789 (0.021)} & {\bf 0.789 (0.028)} & {\bf 0.520 (0.020)} & 9.234 (0.658) \\
\hline
\hline
\vspace{1mm}
Ablated JIMA & NF$^{\mbox{x}}_{\mbox{utb}}$ & 0.718 (0.013) &  &  &  & 3.254 (0.074) \\
\vspace{1mm}
(interactions only) & NF$^{\mbox{x}}_{\mbox{ut}}$ &  & 0.804 (0.022) &  &  & 1.119 (0.056) \\
\vspace{1mm}
& NF$^{\mbox{x}}_{\mbox{ub}}$ &  &  & 0.850 (0.066) &  & 1.095 (0.061) \\
\vspace{1mm}
& NF$^{\mbox{x}}_{\mbox{tb}}$ &  &  &  & 0.551 (0.226) & 0.966 (0.019) \\
\hline
Ablated JIMA & NF$_{\mbox{utb}\oplus\mbox{ut}}$ & 0.709 (0.013) & 0.795 (0.021) &  &  & 4.498 (0.125) \\
 (joint modeling only)& NF$_{\mbox{utb}\oplus\mbox{ut}\oplus\mbox{ub}}$ & 0.708 (0.013) & 0.792 (0.020) & 0.809 (0.025) &  & 6.719 (0.304) \\
& NF$_{\mbox{utb}\oplus\mbox{ut}\oplus\mbox{ub}\oplus\mbox{tb}}$ & 0.705 (0.013) & 0.794 (0.027) & 0.809 (0.024) & 0.547 (0.228) & 9.186 (0.549) \\
\hline
\hline
\end{tabular}
\end{center}
\label{JIM_results-mae}
\end{tiny}
\end{table}

\clearpage

\subsection*{Additional Figures and Tables for the Online Experiment}

\begin{figure*}[htbp]
	\centering
	\captionsetup{width=\linewidth}
	\includegraphics[width=0.8\textwidth]{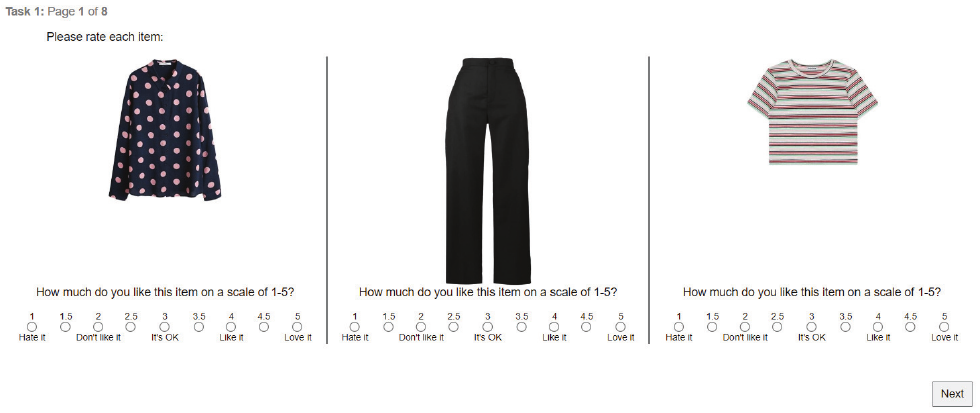}
	\centering
	\caption{\textbf{Example interface for rating the presented individual clothing items}}
	\label{fig:task1}
	\vspace{-1em}
\end{figure*}

An example web page for Task 1 in Section \ref{sec:online} is displayed in Figure \ref{fig:task1}.

\begin{figure*}[htbp]
	\centering
	\captionsetup{width=\linewidth}
	\includegraphics[width=0.6\textwidth]{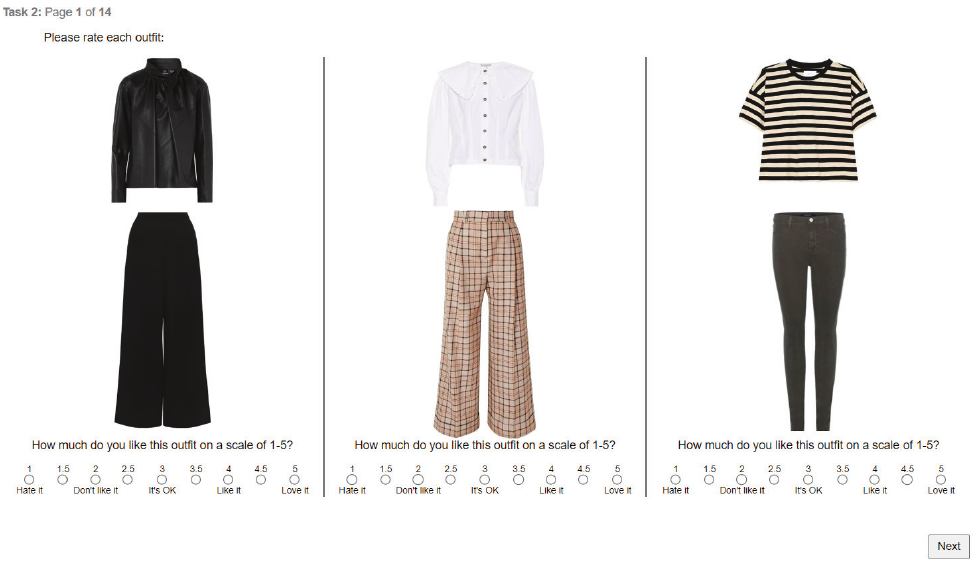}
	\centering
	\caption{\textbf{Example interface for rating the presented outfits}}
	\label{fig:task2}
	\vspace{-1em}
\end{figure*}

An example web page for Task 2 in Section \ref{sec:online} is displayed in Figure \ref{fig:task2}.

The survey items used to measure fashion-related constructs are listed in Table \ref{tab:fashion}
\begin{table}[htbp]
	\centering\caption{\textbf{Survey Items on Fashion and Clothing}}
		\begin{tabular}{p{0.45\textwidth}}
			\hline
			\textbf{Fashion Opinion Leadership}\\ \hline
			When it comes to fashion, I am among the least likely of my friends to be thought of as an advice-giver. \\
			I believe I am a very good source of advice about fashion. \\
			Others consult me for information about the latest fashion trends.\\
			I enjoy discussing fashion.\\ [0.5em]
			\hline
			\textbf{Clothing Interest} \\ \hline
			I would like to be considered one of the best-dressed people. \\
			I like to read and study fashion magazines. \\
			Planning and selecting my wardrobe can be included among my favorite activities. \\
			I enjoy clothes like some people do such things as books, records, and movies.\\
			\hline
	\end{tabular}
    \label{tab:fashion}
	\vspace{-10pt}
\end{table}
\end{document}